\documentclass{emulateapj}

\newcommand{\Msun}{$\mathrm{M_\odot}$}
\newcommand{\Zsun}{$\mathrm{Z_\odot}$}
\newcommand{\Zsmc}{$\mathrm{Z_{SMC}}$}

\shorttitle{SNIb/c progenitors}
\shortauthors{Yoon et al.}

\usepackage{amsmath}

\begin{document}

\title{Type Ib/c supernovae in binary systems~I. \\
Evolution and properties of the progenitor stars}

\author{S.-C. Yoon\altaffilmark{1,2}}

\author{S. E. Woosley\altaffilmark{1}}

\author{N. Langer\altaffilmark{2,3}}


\altaffiltext{1}{UCO/Lick Observatory, Department of Astronomy \& Astrophysics, University of California, Santa Cruz, CA95064}
\altaffiltext{2}{Argelander Institut f\"ur Astronomie, University of Bonn, Auf dem H\"ugel 71,
D-53121, Bonn, Germany}
\altaffiltext{3}{Astronomical Institute, Utrecht University, Princetonplein 5, 3584 CC, Utrecht, The Netherlands}

\begin{abstract}

We investigate the evolution of Type Ib/c supernova (SN Ib/c) progenitors in
close binary systems, using new evolutionary models that include the effects of
rotation, with initial masses of 12 -- 25~\Msun{} for the primary components,
and of single helium stars with initial masses of 2.8 -- 20~\Msun{}.  We find
that, despite the impact of tidal interaction on the rotation of primary stars,
the amount of angular momentum retained in the core at the presupernova stage
in different binary model sequences converge to a value similar to those found
in previous single star models. This amount is large enough to produce
millisecond pulsars, but too small to produce magnetars or long gamma-ray
bursts.  We employ the most up-to-date estimate for the Wolf-Rayet mass loss
rate, and its implications for SN Ib/c progenitors are discussed in detail.  In
terms of stellar structure, SN Ib/c progenitors in binary systems at solar metallicity are predicted
to have a wide range of final masses up to about 7~\Msun, with helium envelopes
of $M_\mathrm{He} \simeq 0.16 - 1.5~\mathrm{M_\odot}$.  Our results
indicate that, if the lack of helium lines in the spectra of SNe Ic were due to
small amounts of helium (e.g. $M_\mathrm{He} \la 0.5$), the distribution
of both initial and final masses of SN Ic progenitors should be bimodal.
Furthermore, we find that a thin hydrogen layer ($0.001~\mathrm{M_\odot} \la
M_\mathrm{H} \la 0.01~\mathrm{M_\odot}$) is expected to be present in
many SN Ib progenitors at the presupernova stage.  We show that the presence of
hydrogen, together with a rather thick helium envelope, can lead to a
significant expansion of some SN Ib/c progenitors by the time of supernova
explosion.  This may have important consequences for the shock break-out and
supernova light curve.  We also argue that some SN progenitors with thin
hydrogen layers produced via Case AB/B transfer might be related to Type IIb
supernova progenitors with relatively small radii of about
$10~\mathrm{R_\odot}$. 

\end{abstract}


\section{Introduction}

It is generally believed that Type Ib and Type Ic supernovae result from core collapse
events of naked helium stars.  The helium stars are thought to be produced by
the loss of the hydrogen envelope, via stellar winds mass loss from massive single
stars or via mass transfers in close binary systems.  

According to recent stellar models adopting the most up-to-date stellar winds
mass loss rates \citep{Meynet03, Meynet05, Eldridge06, Limongi06, Georgy09},
the final masses ($M_\mathrm{f}$) of helium stars produced by mass-losing
single stars appear to be too massive to produce typical SNe Ib/c (i.e.,
$M_\mathrm{f} > 10$~\Msun{} at solar metallicity).  Although the limiting mass
for BH formation is not yet well determined, given their high binding energy,
such massive progenitors of $M_\mathrm{f} > 10$~\Msun{} are likely to form
black holes (BHs), producing faint supernovae or no supernova at
all~\citep[cf.][]{Fryer99}.  Although very bright SNe Ib/c like SN 1998bw could
be produced from such  massive helium stars if, for example, powered by rapid
rotation \citep[e.g.,][]{Woosley93, Burrows07}, such events are shown to be
rare \citep[e.g.,][]{Podsiadlowski04, Guetta07}.  

By contrast, helium stars with a wide range of  masses ($2.0~\mathrm{M_\odot}
\la M_\mathrm{He} \la 25~\mathrm{M_\odot}$) can be made from $12...60$~\Msun{}
primary components in close binary systems via the so-called Case A/B mass
transfer.\footnote{Case A, B or C mass transfer denotes mass transfer from
the primary star during core hydrogen burning, helium core
contraction/beginning of core helium burning, or    core helium burning and
later stages, respectively. On the other hand, if mass transfer occurs during
helium core contraction/beginning of core helium burning from a star that has
already undergone Case A mass transfer, such a mass transfer phase is called
Case AB. Case ABB or Case BB mass transfer denotes  mass transfer from the
primary star during core helium burning and/or later stages, which has already
undergone Case AB or Case B mass transfer, respectively.} Many of them may end
their life as bright SNe Ib/c leaving neutron stars as remnants, if their final
masses are less than about 7 -- 10~\Msun{}.   Population studies indeed show that
close binary stars can produce a sufficient number of SNe Ibc to explain their
observed rate, without the need of invoking single star progenitors
\citep[e.g.,][]{Podsiadlowski92, deDonder98, Eldridge08}.  Therefore, it is
most likely that the majority of typical SNe Ibc are produced in binary
systems.

The observational evidence for the connection between SNe Ibc and long
gamma-ray bursts (GRBs) has particularly motivated many observational studies
to better understand SNe Ibc since the last decade \citep[see][for a
review]{WB06}. Theoretical stellar models of SNe Ibc progenitors are thus
highly required nowadays.  The most comprehensive studies on the detailed
characteristics of SNe Ibc progenitors in binary systems were conducted by
\citet{Woosley95} (hereafter, WLW95) using mass-losing pure helium star models,
and by \citet{Wellstein99} (hereafter, WL99) using self-consistent binary star
models.  
Although more recent theoretical studies on  SNe Ibc progenitors
in binary systems can be found in the literature, they have been focused on
long GRB progenitors or stellar populations, rather than on the detailed nature
of typical SNe Ibc progenitors \citep[e.g.,][]{Brown00, Izzard04, Petrovic05a,
Cantiello07, Heuvel07, Detmers08, Eldridge08}. 

In this paper, we revisit the problem of SNe Ibc progenitors in close binary
systems using both binary star and single helium star models up to the neon burning
stage, with updated physics of two important ingredients. One is rotation,
which was not considered in WLW95 and WL99, and the other is the mass loss rate
of Wolf-Rayet (WR) stars (Sect.~\ref{sect:review}).

This paper is organized as follows.  In Sect.~\ref{sect:review}, we  briefly
review recent developments of stellar evolution models regarding the effects of
rotation and the WR star mass loss rate, arguing for the need of updated physics
in binary star models.  Our adopted physical assumptions and numerical method
are discussed in Sect.~\ref{sect:method}.  In the following section
(Sec.~\ref{sect:rotation}), using our binary star evolution models including
the effect of rotation and the transport of angular momentum due to
hydrodynamic instabilities and magnetic torques, we explore the role of tidal
interaction and mass transfer in the redistribution of angular momentum in
primary stars.  In Sec.~\ref{sect:sn}, the nature of SNe Ibc progenitors is
investigated in terms of final masses, masses of helium and hydrogen layers,
radii and mass loss rates at the presupernova stage, assuming these properties
do not significantly change from neon burning to core collapse.  For this
purpose, we also present mass-losing single helium star models as a complement
to our binary star models, given that the parameter space explored with our
binary model sequences is limited.  We conclude the paper by discussing
observational implications of our results, in Sect.~\ref{sect:discussion}.

\section{Rotation and Wolf-Rayet winds}\label{sect:review}

\subsection{Rotation}

Rotation has particular roles in the evolution  massive stars as it changes
the stellar structure, induces chemical mixing and enhances mass loss due to
stellar winds \citep{Maeder00, Heger00a}.  During the last decade, several
authors have calculated massive star models up to the pre-supernova stage,
considering the redistribution of angular momentum and chemical species due to
rotationally induced hydrodynamic instabilities, such as Eddington-Sweet
circulations, and the shear instability \citep{Heger00a, Hirschi04}. Although these
models could explain some observational aspects such as surface abundances of
CNO elements of massive stars and Wolf-Rayet (WR) star populations at different
metallicities \citep[e.g.,][]{Maeder00, Heger00b, Meynet05}, their adopted
angular momentum transport mechanisms turned out to be too inefficient to explain
the observed spin rates of stellar remnants \citep{Suijs08}.  I.e., their models predict nearly
two orders of magnitude higher spin rates of white dwarfs and young neutron
stars than the observed ones. These models also imply that almost all WR
stars can retain enough angular momentum to produce GRBs, either by magnetar or
collapsar formation depending on the final mass.  This gives a nearly 1000 times
higher ratio of GRBs to SNe than the observationally implied value.  On the
other hand, \citet{Spruit99, Spruit02} suggested that magnetic torques resulting
from dynamo actions in differentially rotating radiative layers (the so-called
Spruit-Tayler dynamo) should be the dominant angular momentum transport mechanism compared to the pure
hydrodynamic instabilities.  Recent magnetic models that adopt the
Spruit-Tayler dynamo according to the prescription by \citet{Spruit02} are
indeed more consistent with observations in terms of the spin rates
of stellar remnants \citep{Heger05, Suijs08}.  Magnetic models also better
explain the fact that long GRBs are rare events compared to normal core
collapse supernovae.  Although the Spruit-Tayler dynamo mechanism is still
subject to many uncertainties \citep{Denissenkov07, Zahn07}, these recent
studies indicate that an efficient angular momentum transport mechanism comparable to what the
Spruit-Tayler dynamo predicts is needed to understand the observations.

The above discussion is based on single star models, and the role of rotation
in the evolution of massive binary stars remained relatively unexplored.
Massive stars in close binary systems are supposed to experience an exchange of
mass and angular momentum via mass transfer and tidal interaction, and thus
the evolution of binary stars is more complex than that of single
stars. \citet{Wellstein01} and \citet{Langer03} presented, for the first time,
self-consistent calculations of massive binary star evolution models
including many relevant effects of rotation and binary interactions: the change of
the stellar structure due to the centrifugal force, transport of angular momentum
and chemical species due to rotationally induced hydrodynamic instabilities,
tidal interaction, transfer of mass from the primary, accretion of mass and
angular momentum of the secondary, and resulting changes of the orbit. Their
work indicates that the secondary can be easily spun up by mass accretion even
up to critical rotation, thus modulating the mass accretion efficiency by
the interplay of the enhanced mass loss due to rotation from the spun-up
secondary and the mass transfer from the primary.  This effect provides
important clues to better understand the evolutionary paths of some observed
X-ray and WR star binary systems as discussed by \citet{Langer03} and
\citet{Petrovic05b}.  Their non-magnetic models also show that massive stars
may end up with different core spin rates depending on
the history of mass loss/gain during binary evolution, that could be related to the
observational diversity of core-collapse supernovae.  More recently,
\citet{Petrovic05a} and \citet{Cantiello07} included the Spruit-Tayler dynamo
in their binary models, and discussed possible evolutionary paths of massive
binary stars towards long GRBs.  In this paper, we present new evolutionary
calculations of magnetic (i.e., the Spruit-Tayler dynamo is included) massive
binary stars for a large parameter space,  
focusing on the evolution of the primary stars to investigate
the nature of typical SNe Ibc progenitors.

\subsection{Mass loss due to Wolf-Rayet winds}\label{sect:wrwind}

\begin{figure}
\epsscale{1.}
\plotone{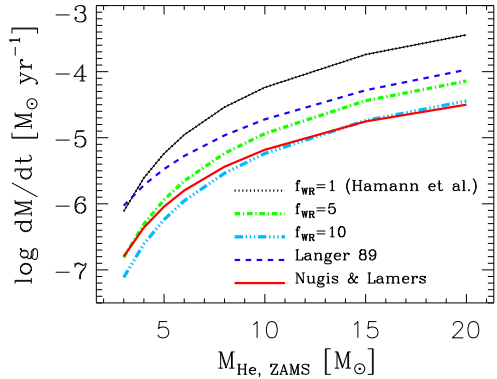}
\caption{
Wolf-Rayet mass loss rates from helium stars on the zero-age main sequence at solar metallicity. 
The mass loss rates of Hamann et al. (see Eq.~\ref{eq1}), \citet{Langer89} and \citet{Nugis00} are given by
dotted, dashed and solid lines, respectively. The dotted-dashed and three dotted-dashed lines denote
the Hamann et al. rates divided by a factor of 5 and 10, respectively (i.e., $f_\mathrm{WR} =$ 5 and
10 in Eq.~\ref{eq1}.).  
}\label{fig:wrwind}
\end{figure}

WLW95 employed the mass-dependent mass loss rate of WR stars of \citet{Langer89}
for their helium star models, and WL99 used those of \citet{Hamann82} and
\citet{Hamann95}.  Both  studies concluded that the final masses of SNe
Ibc progenitors in binary systems should converge to a limited range of 2.2 --
3.6~\Msun, even for an initial helium star mass of 20~\Msun.  Later
developments of the WR wind theory considering clumpies 
pointed out, however, that the WR mass loss rates
used by WLW95 and WL99 are significantly overestimated \citep{Hamann98,
Nugis00}.  For example, Fig.~\ref{fig:wrwind} shows that, for helium stars on
the zero age main sequence, the mass loss rate given by \citet{Nugis00} is
almost an order-of-magnitude lower than that of Hamann et al. used in WL99. The
most recent theoretical models of WR winds by \citet{Vink05} and
\citet{Graefener08}  also give WR mass loss rates compatible to the Nugis \&
Lamers rate \citep[cf. Figure 1 in][]{Yoon05}.   

The reduced WR mass loss rates have been considered in many recent massive
single star models \citep[e.g.,][]{Meynet03, Meynet05, Eldridge06}, suggesting
large final masses of WR stars at the presupernova stage ($M_\mathrm{f} >
10$~\Msun{} at solar metallicity) compared to $\sim 4$~\Msun{} found by
\citet{WLW93}. As argued in the introduction, this leads to the interesting
conclusion that most core collapse events occurring in single WR stars should
not produce bright supernovae, rendering the binary star channel even more
important for SNe Ibc.  Surprisingly, the effect of the new WR mass loss rate
on SNe Ibc progenitors in close binary systems has not been much discussed or
largely overlooked.  To our knowledge, \citet{Pols02} are the only authors who
addressed this issue in detail.  Although WL99 also presented some model
sequences with a reduced WR mass loss rate, their adopted reduction factor was
rather modest (2 times smaller than the Hamann et al. rate). Here we consider
the Hamman et al. WR mass loss rate reduced by factors of 5 and 10, which
reflects the most recent result as discussed above.

\section{Numerical method and physical assumptions}\label{sect:method}

We have used the same stellar evolution code as in \citet{Cantiello07}, that
follows the simultaneous evolution of the two stellar components of a binary
system.  The effect of the centrifugal force on stellar structure is considered
following \citet{Endal76}. The transport of angular momentum and chemical
elements is treated as diffusion considering Eddington Sweet circulations, the
shear instability, the Goldreich-Schubert-Fricke instability and the
Spruit-Tayler dynamo, as described in \citet{Petrovic05a}. We use
$\alpha_\mathrm{SEMI} = 1.0$ for the semi-convection parameter
\citep{Langer83}, as in \citet{Yoon06}.  

The stellar winds mass loss rate is determined following \citep{Kudritzki89} with a
metallicity scaling of $(Z/Z_\odot)^{0.69}$ \citep{Vink01} for the main sequence phase.  
WR wind mass loss rates are computed according to Hamman et al. with a
correction factor $f_\mathrm{WR}$, 
and with a metallicity
dependence of $\dot{M} \propto (Z_\mathrm{init}/Z_\odot)^{0.86}$ \citep{Vink05}:
\begin{eqnarray}\label{eq1}
\log\left(\frac{\dot{M}_\mathrm{WR}}{\mathrm{M_\odot~yr^{-1}}}\right)  = 
  & -11.95 + 1.5\log L/\mathrm{L_\odot} - 2.85 X_\mathrm{s} \nonumber    \\
  &  + 0.86\log(Z_\mathrm{init}/Z_\odot)-\log f_\mathrm{WR} \nonumber \\
  &   \mathrm{for~~ \log L/\mathrm{L_\odot} > 4.5,}  \\
  =  & -35.8 + 6.8\log L/\mathrm{L_\odot} - 2.85 X_\mathrm{s}  \nonumber  \\
   & + 0.86\log(Z_\mathrm{init}/Z_\odot) -\log f_\mathrm{WR} \nonumber \\
   & \mathrm{for~~\log L/\mathrm{L_\odot} \le 4.5}~~. \nonumber
\end{eqnarray}
We use $f_\mathrm{WR} = 5$ or 10 in most model sequences, which means the WR mass loss rate
by Hamman et al. is lowered 5 or 10 times, to consider the most recent
estimates (see Fig.~\ref{fig:wrwind}).  The enhancement of
stellar winds mass loss due to the centrifugal force is considered by
\begin{equation}\label{eq2} \frac{\dot{M}}{\dot{M}(v_\mathrm{rot}=0)} = \min
\left [ \left ( \frac{1}{1-\Omega} \right )^{0.43}, 0.5
\frac{M}{\tau_\mathrm{KH}} \right] ~~, \end{equation} where $\Omega =
v_\mathrm{rot}/v_\mathrm{crit}$ with $v_\mathrm{crit} = \sqrt{GM(1-\Gamma)/R}$
and $\Gamma$ is the Eddington factor \citep{Langer98}.  Here, to prevent a
singularity that may occur as $v_\mathrm{rot}$ approaches $v_\mathrm{crit}$,
the mass loss rate is limited to $\dot{M} \le 0.5 M/\tau_\mathrm{KH}$,
where $\tau_\mathrm{KH}$ is the thermal time scale of the star.

The binary orbit is assumed to be circular, and the Roche lobe radius is
determined according to the approximation of \citet{Eggleton83}.  The mass loss
rate of the Roche-lobe filling component through the first Lagrangian point is
implicitly computed using the method given by \citet{Ritter88}.  The equation
of motion of a test particle is numerically solved to calculate the amount of
angular momentum of the accreted matter if the transfered matter directly hits
the secondary star, and the Keplerian value is assumed otherwise
\citep{Wellstein01}.  The change of the orbital period due to mass transfer and
stellar wind mass loss is considered according to \citet{Podsiadlowski92}.  We
follow \citet{Brookshaw93} to determine the amount of the specific angular
momentum carried away from the orbit by stellar winds.  

Tidal synchronization is considered following \citet{Wellstein01} \citep[see also][]{Detmers08}.  
We assume a synchronization time scale
according to \citet{Tassoul87, Tassoul00} who considered
tidally driven meridional circulations as the main mechanism for tidal dissipation:
\begin{align}\label{eq3}
\tau_\mathrm{sync} (yr)  &  = f_\mathrm{sync}
 \frac{1.44 \times 10^{1.6}}{q(1+q)^{3/8}} 
 \left (\frac{L_\odot}{L} \right )^{1/4} \\
& \times \left (\frac{M_\odot}{M} \right )^{1/8}%
\left (\frac{R}{R_\odot} \right )^{9/8} \left ( \frac{d}{R} \right )^{33/8}~~ \nonumber, 
\end{align}
where $q$ denotes the mass ratio and $d$ the orbital separation. 
This prescription gives a much shorter time scale than that given by \citet{Zahn77}.
Given that the physics of tidal dissipation is much debated in the literature
~\citep{Langer09}, we introduce a parameter $f_\mathrm{sync}$  to
investigate how an extremely fast/slow synchronization may influence the
results.  In most cases, however, we use $f_\mathrm{sync} = 1$. 

A few sequences are also computed with $\tau_\mathrm{sync}$ of \citet{Zahn77} for comparison:
\begin{align}\label{eq4}
\frac{1}{\tau_\mathrm{sync}} & =  5\left (\frac{GM}{R^3} \right )^{1/2} q^2 (1+q)^{5/6}  \nonumber\\
           & \times \frac{MR^2}{I} E_2 \left (\frac{R}{d} \right )^{17/2}~~, 
\end{align}
where $I$ is the moment of the star, and $E_2$ a constant measuring the
coupling between the tidal potential and the gravity mode. Using the data of
Table~1 in \citet{Zahn77}, we constructed a fitting formula for $E_2$ as the
following:  \begin{equation}\label{eq5} E_2  = 10^{-1.37} \left
(\frac{R_\mathrm{conv}}{R} \right)^8~~, \end{equation} where $R_\mathrm{conv}$
is the radius of the convective core.  Note that both prescriptions by Tassoul
and Zahn are not appropriate for a star with a convective envelope
\footnote{On the other hand, note that a recent study by \citet{Toledano07} suggests that  
intermediate mass main sequence stars follow the Zahn's synchronization time scale for convective stars.}. 
However, the role of tidal
synchronization is significant only on the main sequence, and not important in
late evolutionary stages as discussed below. 

We computed 45 model sequences for initial masses of the primary star mostly
from 12 to 25 $\mathrm{M}_\odot$ at two different metallicities ($Z=$ 0.02 and
0.004),  for different mass ratios, initial orbital periods, and WR mass loss
rates, as summarized in Table.~\ref{tab1}.  The initial rotational velocity at the equatorial surface of each star
is set to be 20\% of the Keplerivan value. We could not calculate more massive
systems because of a numerical difficulty encountered during the mass transfer
phases, except for Seq.~26 where a primary star of 60~\Msun{} is considered
with a rather large WR mass loss rate (i.e., $f_\mathrm{WR} = 3$).  The adopted
initial orbital periods corresponds
either to Case~A or to Case~B mass transfer.  In the present study, we do not
consider Case~C systems, but briefly discuss the possible outcomes of Case~C
mass transfer in Sect.~\ref{sect:dischydrogen}.  The evolution of the primary
stars is followed up to neon burning in most cases. 

We also present non-rotating single helium star models to discuss SNe Ibc
progenitors in binary systems with initial masses larger than 25~\Msun{}, and
also to compare them with binary star models (Sect.~\ref{sect:sn}).

\section{Redistribution of angular momentum in primary stars}\label{sect:rotation}

In this section, we focus our discussion on the evolution of primary stars and
investigate whether binary evolution via Case~A or Case~B mass transfer could
lead to diverse pre-collapse conditions of SNe Ibc in terms of the amount of
core angular momentum. Although the evolution of mass-accreting secondary stars
is a matter of extreme interest as discussed in \citet{Braun95},
\citet{Petrovic05a} and \citet{Cantiello07},  it is beyond the scope of this
paper.

Here, we first present some results including the Spruit-Tayler dynamo with our fiducial assumption on synchronization time (i.e., $f_\mathrm{sync}=1$), showing that the final amount of angular momentum in the core of the primary star 
is not much affected by different histories of mass loss (i.e., Case AB or Case B; Sect.~\ref{sect:fiducial}).  Then, we discuss the influences
of different assumptions on tidal synchronization and transport process of angular momentum (Sect.~\ref{sect:nonfiducial}).

\begin{figure}
\epsscale{1.}
\plotone{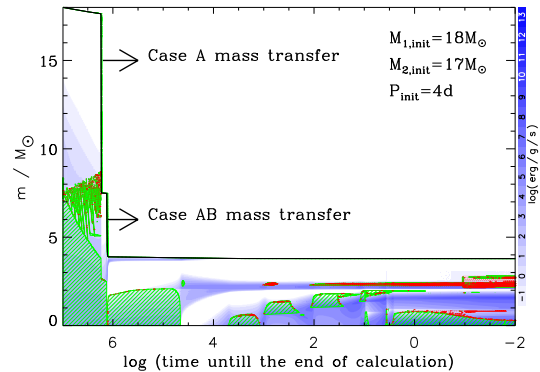}
\caption{Evolution of the internal structure of the primary star in Seq.~14 ($M_\mathrm{1,init}=18~\mathrm{M_\odot}, 
M_\mathrm{2,init}=17~\mathrm{M_\odot}$ and $P_\mathrm{init} = 4~\mathrm{day}$) from ZAMS to the neon burning phase. 
The hatched lines and the red dots denote convective layers and semi-convective layers, respectively.
The different shades give the nuclear energy generation rate, for which the scale is shown on the right hand side.
The surface of the star is marked by the topmost solid line. 
}\label{fig:kippseq9}
\end{figure}

\begin{figure}
\epsscale{1.0}
\plotone{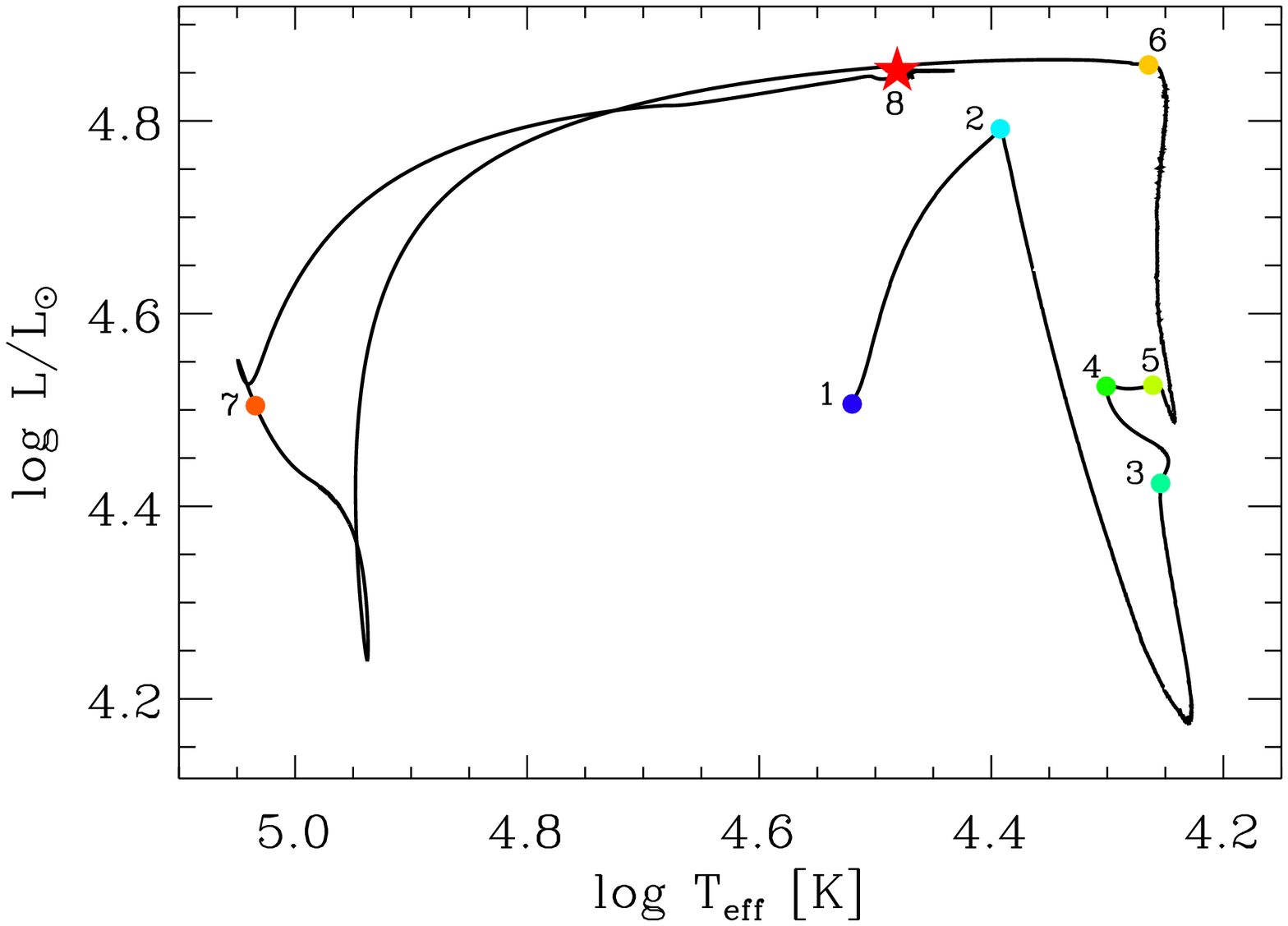}
\plotone{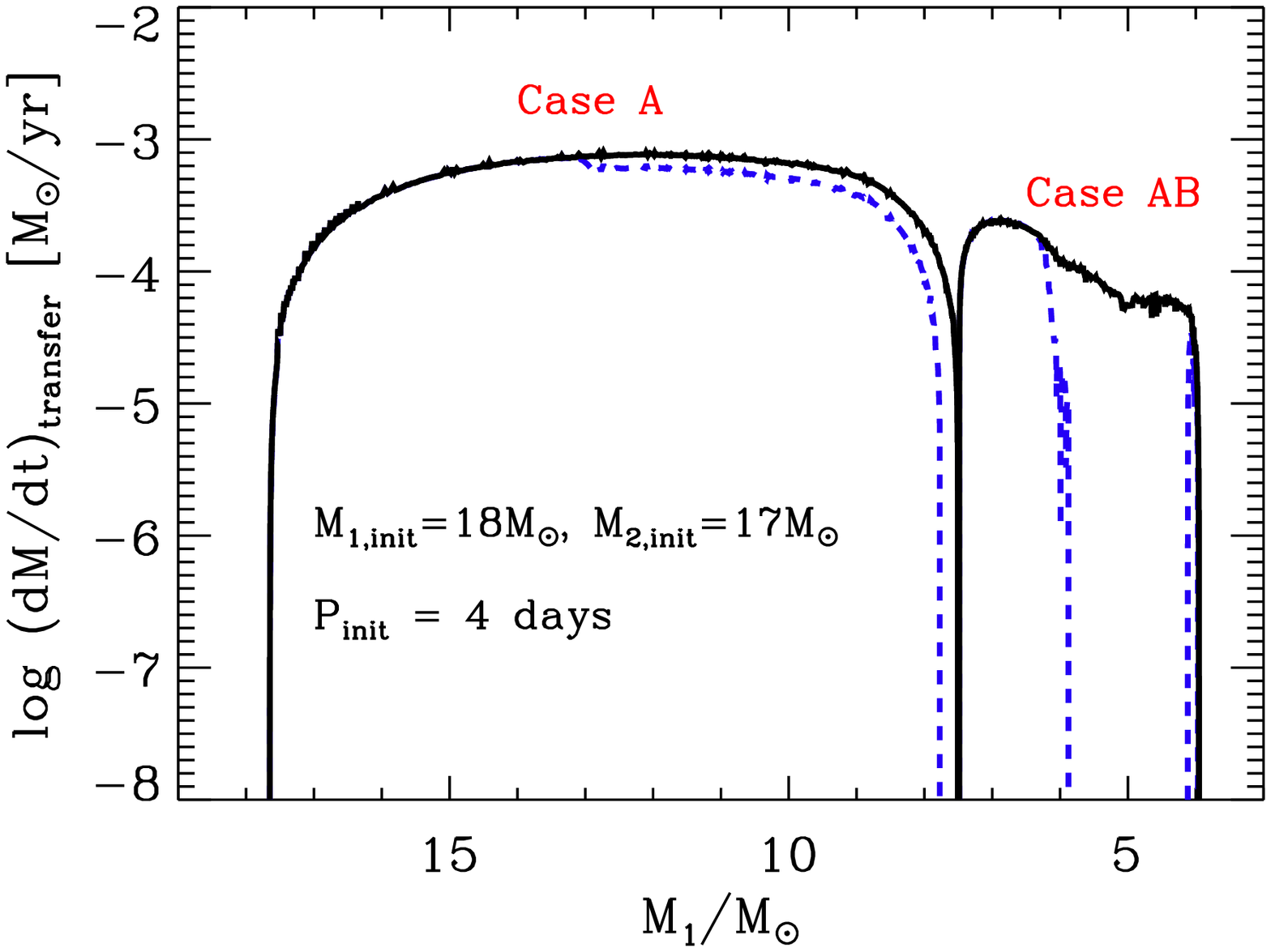}
\caption{\emph{Upper panel}: 
Evolutionary track of the primary star in Seq.~14
($M_\mathrm{1,init}=18~\mathrm{M_\odot}, M_\mathrm{2,init}=17~\mathrm{M_\odot}$
and $P_\mathrm{init} = 4~\mathrm{day}$) in the HR diagram.  The filled circles
on the track mark different evolutionary epochs as the following. 1: ZAMS, 2:
beginning of the Case A mass transfer, 3: end of the Case A mass transfer, 4:
core hydrogen exhaustion, 5: beginning of the Case AB mass transfer, 6: end of
the Case AB mass transfer, 7: core helium exhaustion, 8: neon burning (end of
calculation).  \emph{Lower panel}: Mass transfer rates from the primary star (solid line) 
and mass accretion rates onto the secondary star (dashed line) during the Case A and AB
transfers as a function of the primary star mass.}\label{fig:hrseq9}
\end{figure}

\begin{figure}
\epsscale{1.}
\plotone{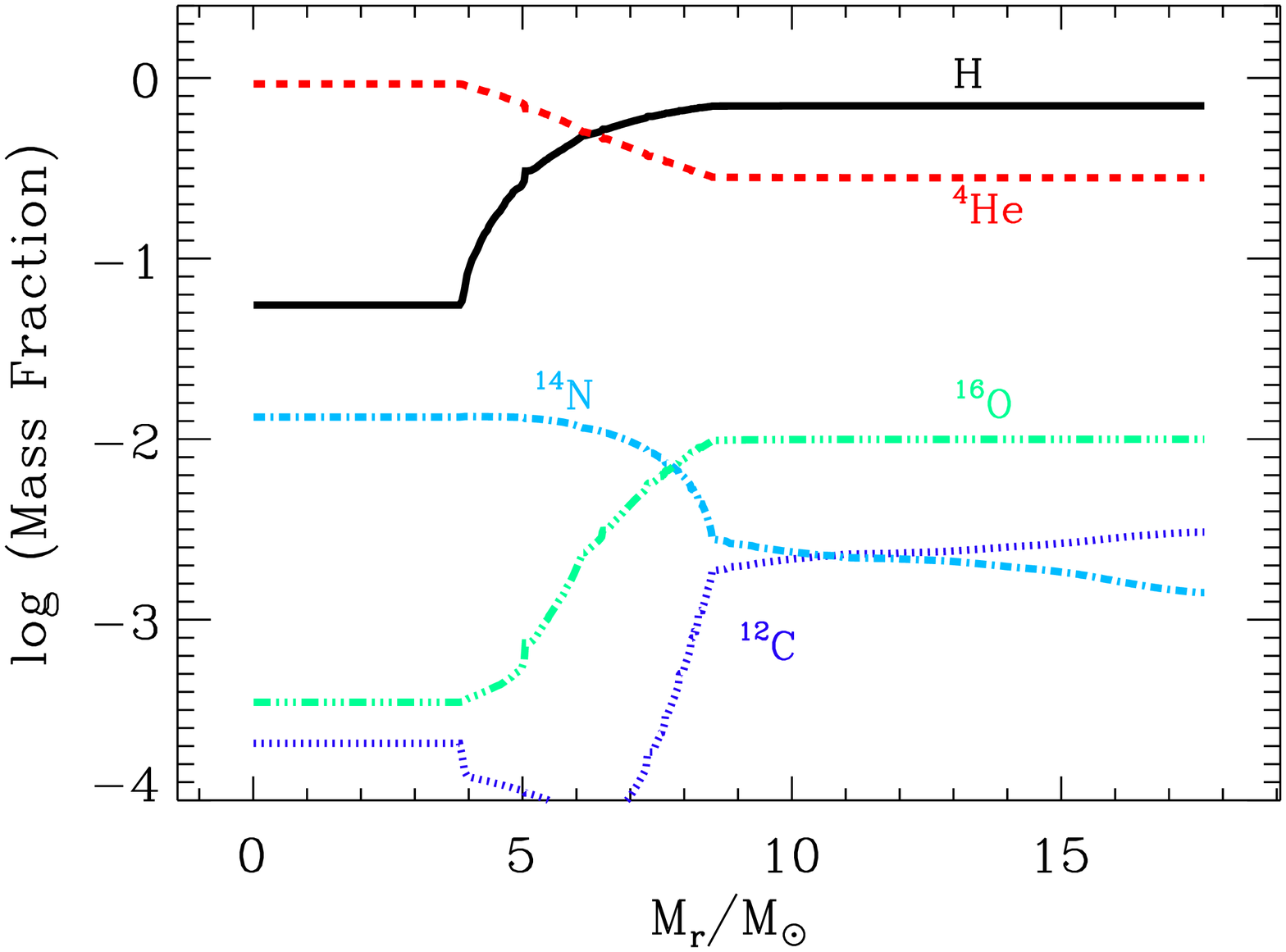}
\plotone{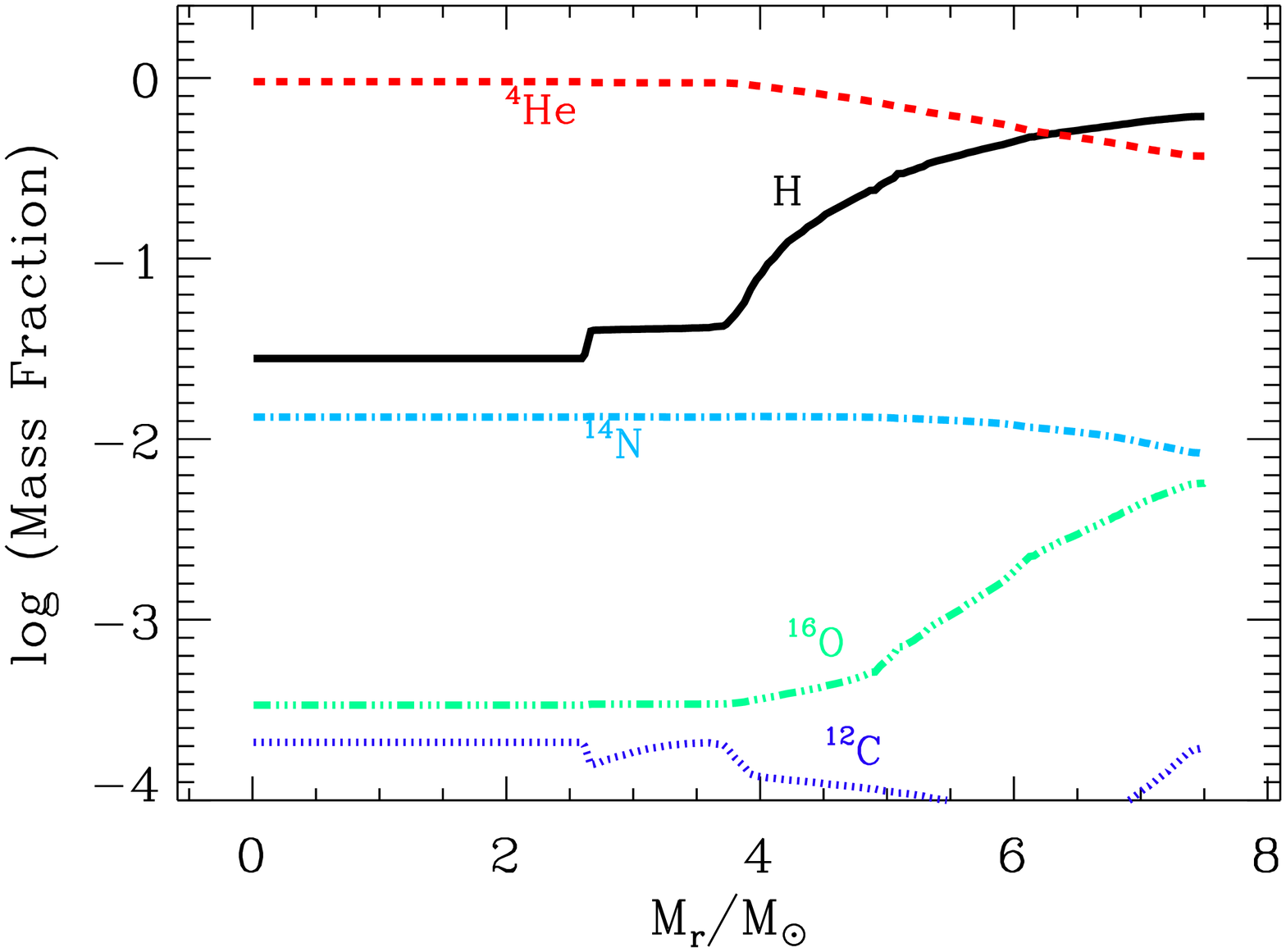}
\plotone{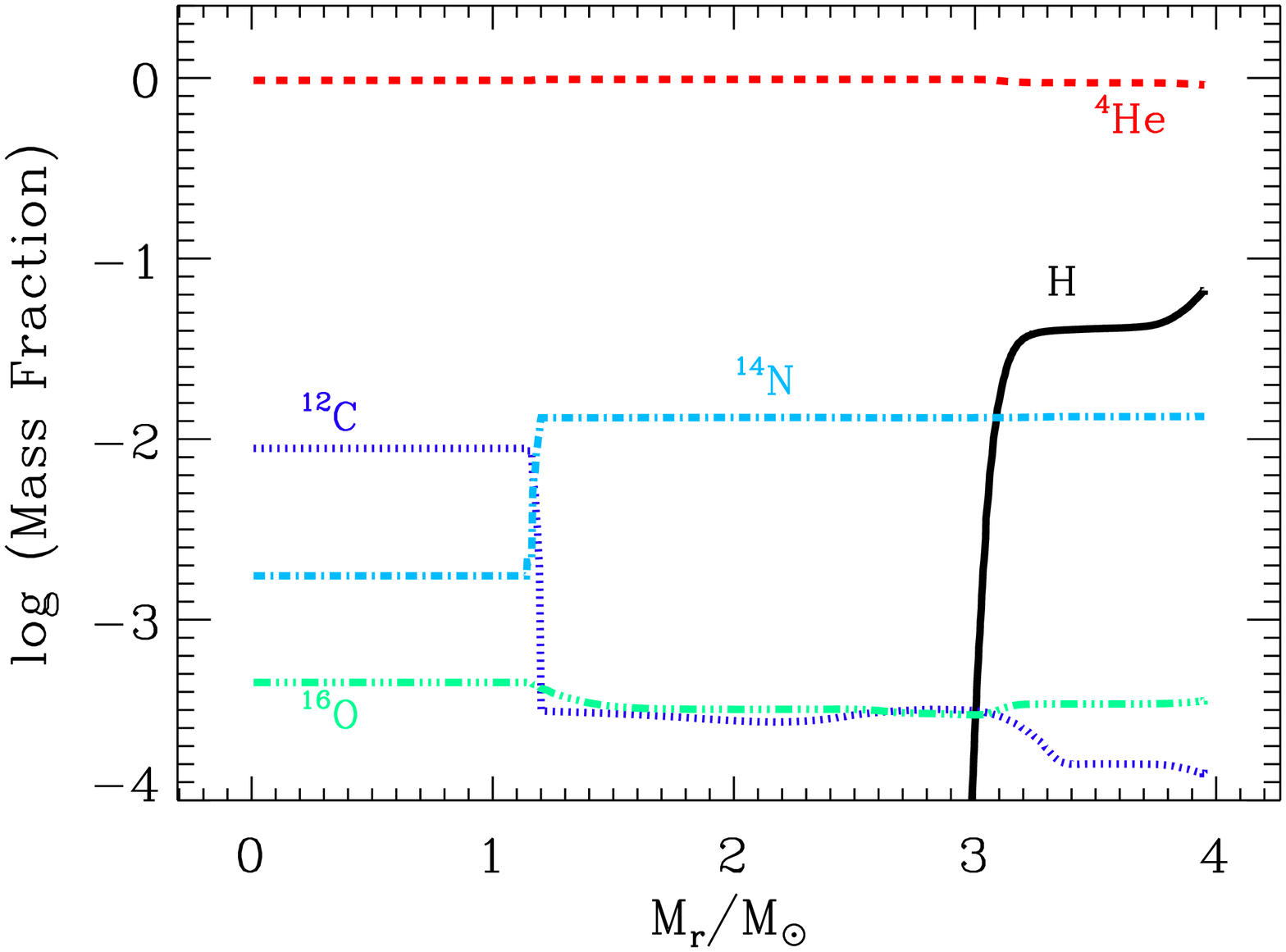}
\plotone{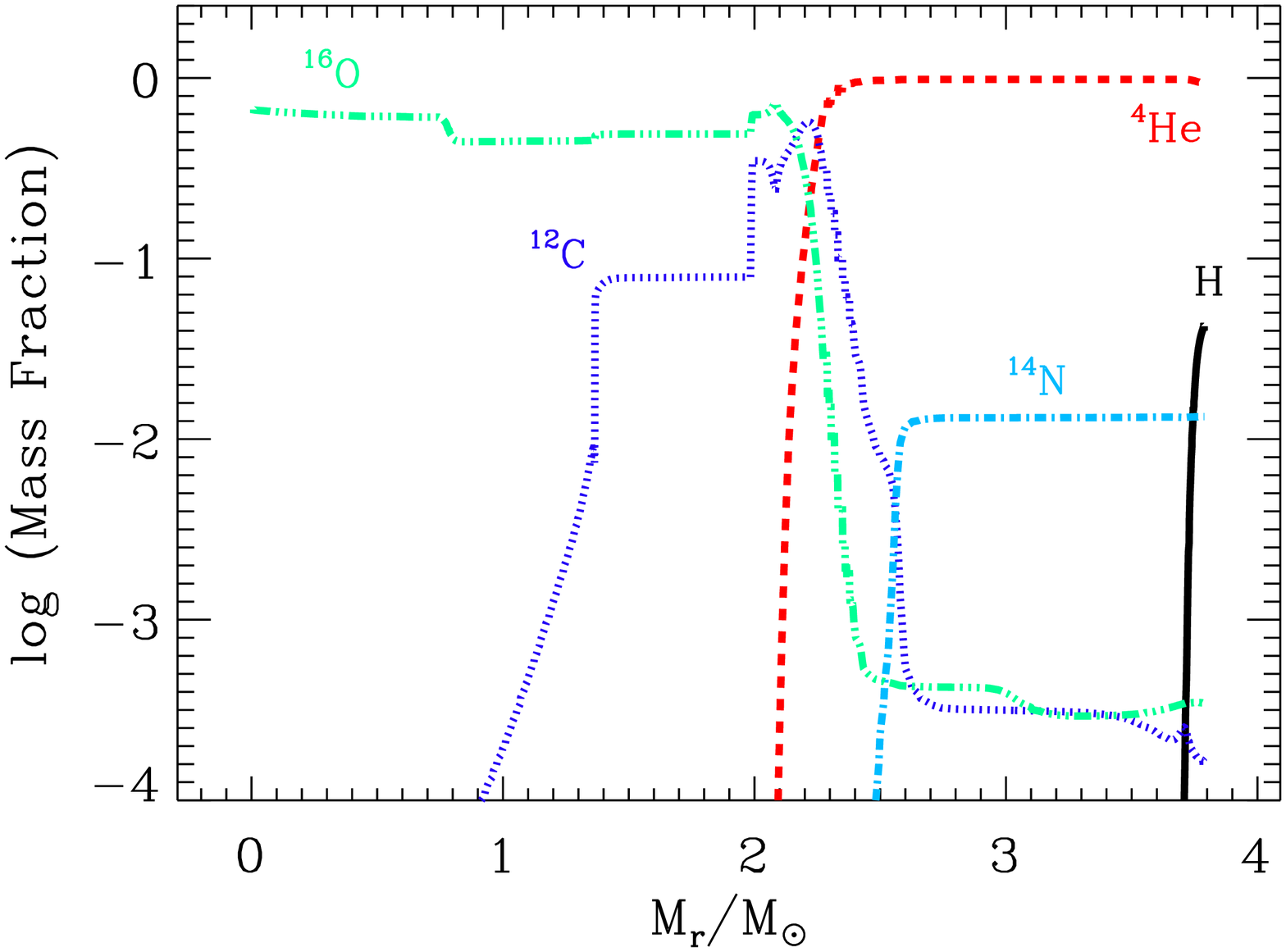}
\caption{
Chemical composition of the primary star in Seq.~14 as a function of 
the mass coordinate, at different evolutionary
epochs. 
\emph{First Panel}: Core H burning (right before the Case A mass  transfer phase)
\emph{Second Panel}: Core H burning (right after the Case A mass  transfer phase)
\emph{Third Panel}: Helium burning (right after the Case AB mass transfer phase) 
\emph{Last Panel}: Neon burning
}\label{fig:chem} 
\end{figure}

\begin{figure}
\epsscale{1.0}
\plotone{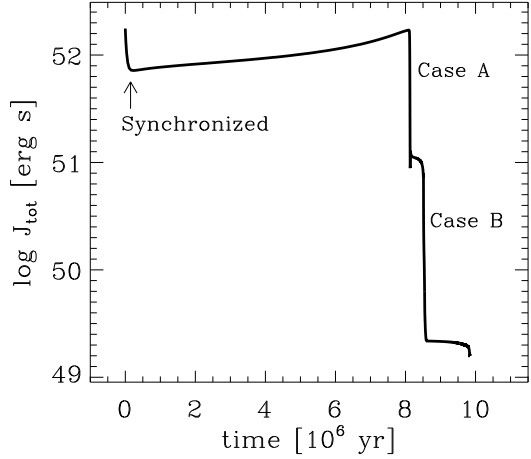}
\caption{
Evolution of the total angular momentum
of the primary star in Seq.~14, as a function of time, 
from the ZAMS until neon burning.  
}\label{fig:totjsync}
\end{figure}

\begin{figure}
\epsscale{1.0}
\plotone{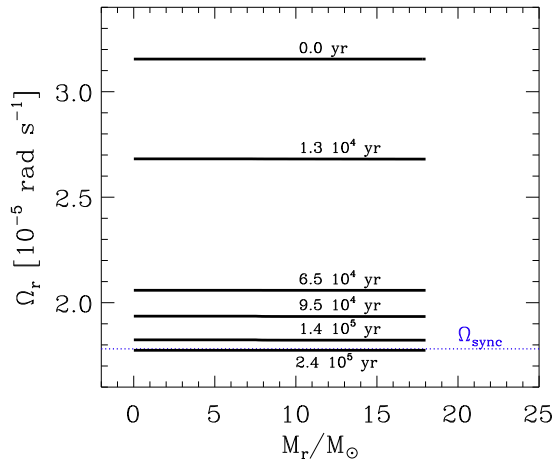}
\caption{
Angular velocity profile in the primary star of Seq.~14 as a function of the mass
coordinate, at 6 different epochs from the ZAMS until $2.4\times10^5$ yr.  The
thin dotted line denotes the angular velocity of the orbit (i.e.,
$\Omega_\mathrm{sync} := 2\pi / P_\mathrm{orbit}$) at $2.4\times10^5$ yr. 
}\label{fig:omegasync}
\end{figure}

\begin{figure}
\epsscale{1.0}
\plotone{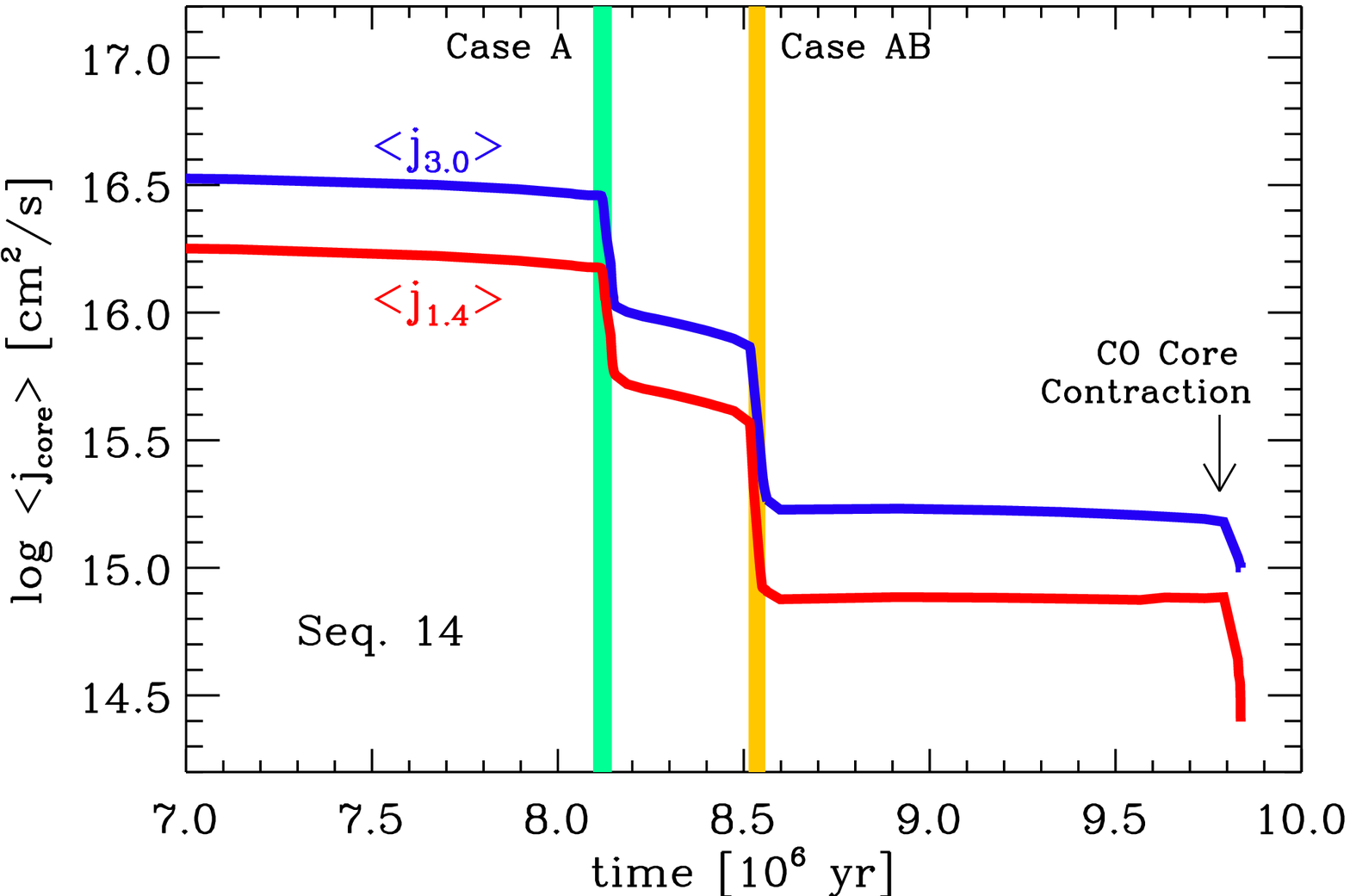}
\plotone{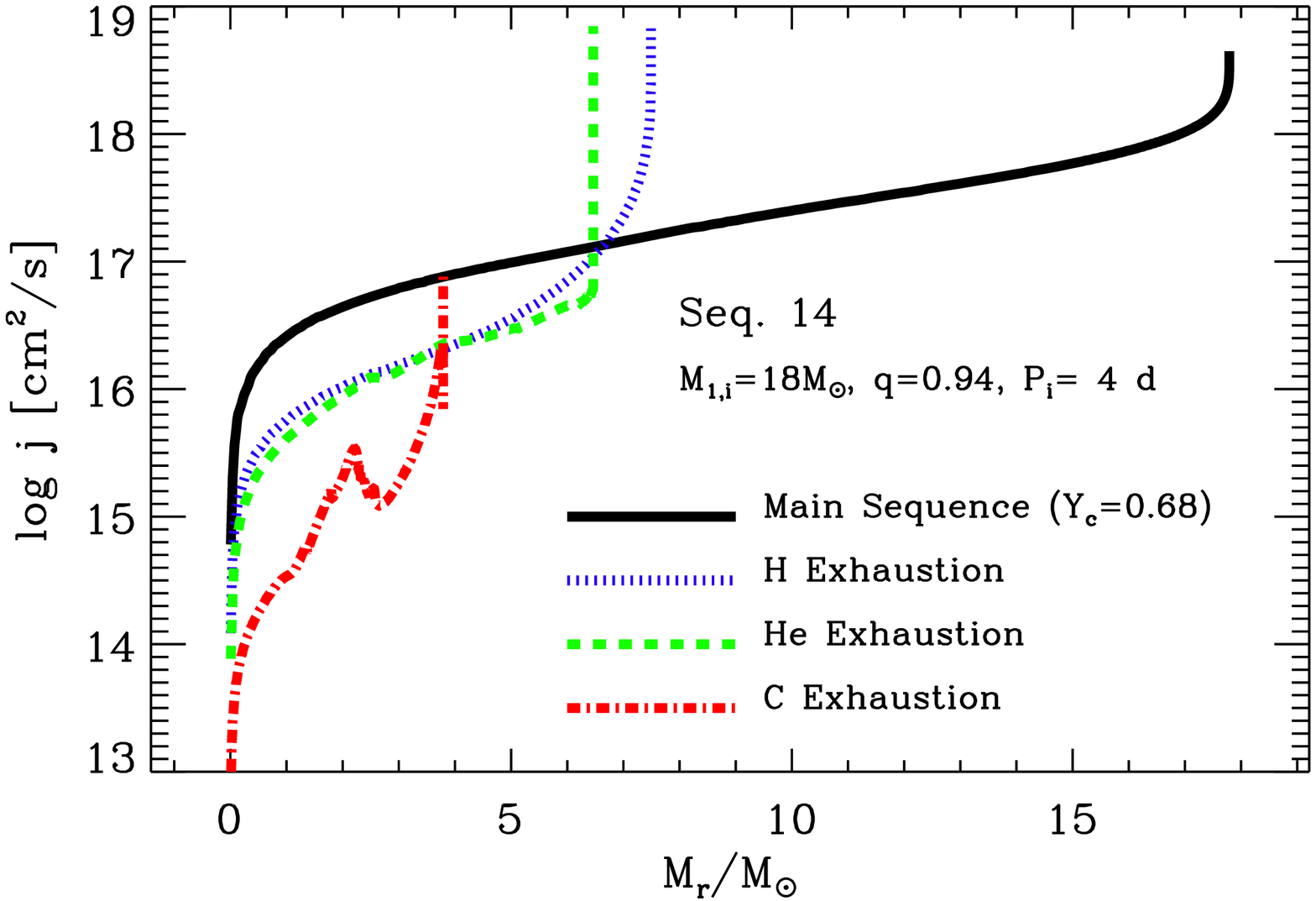}
\plotone{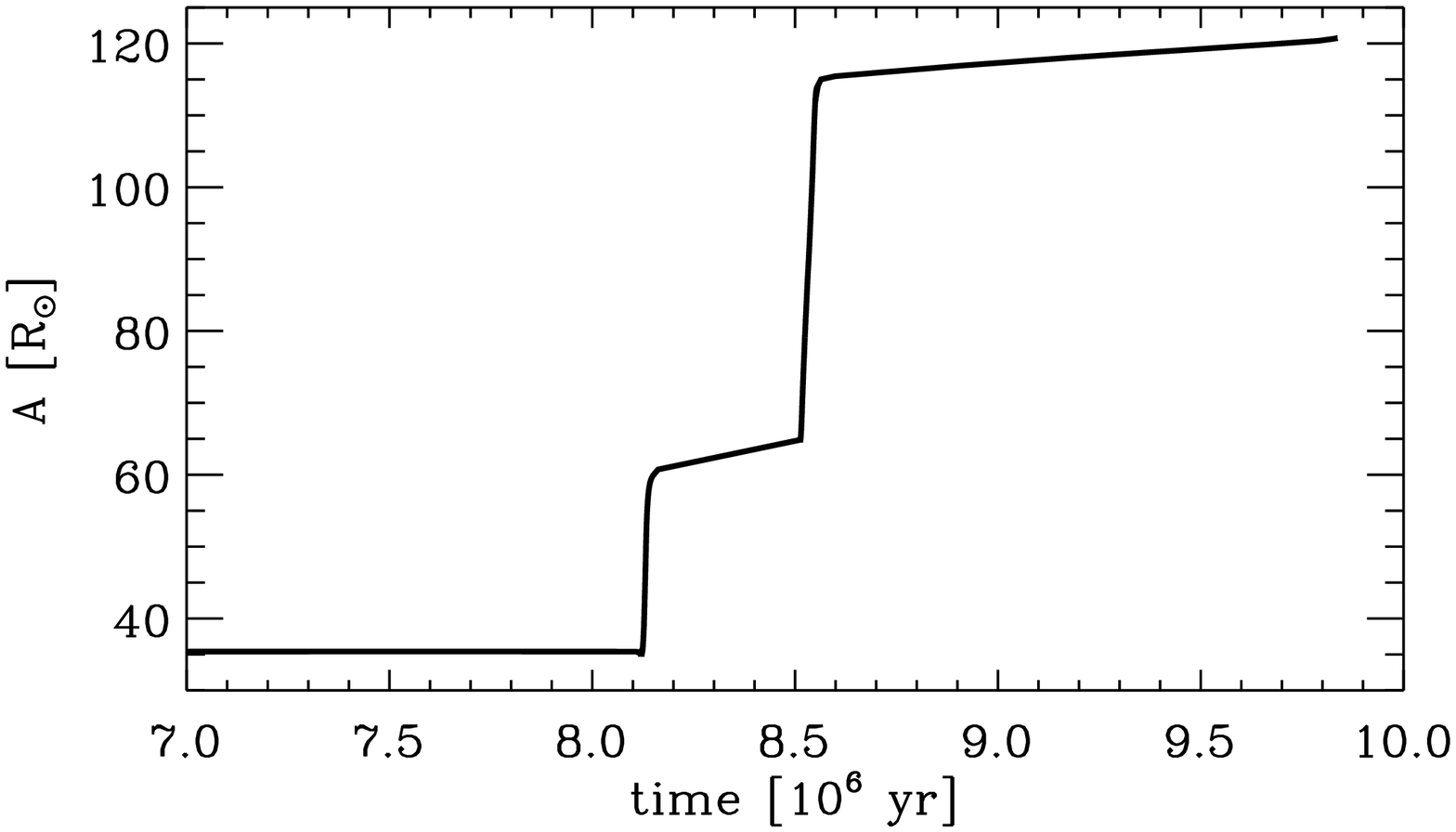}
\plotone{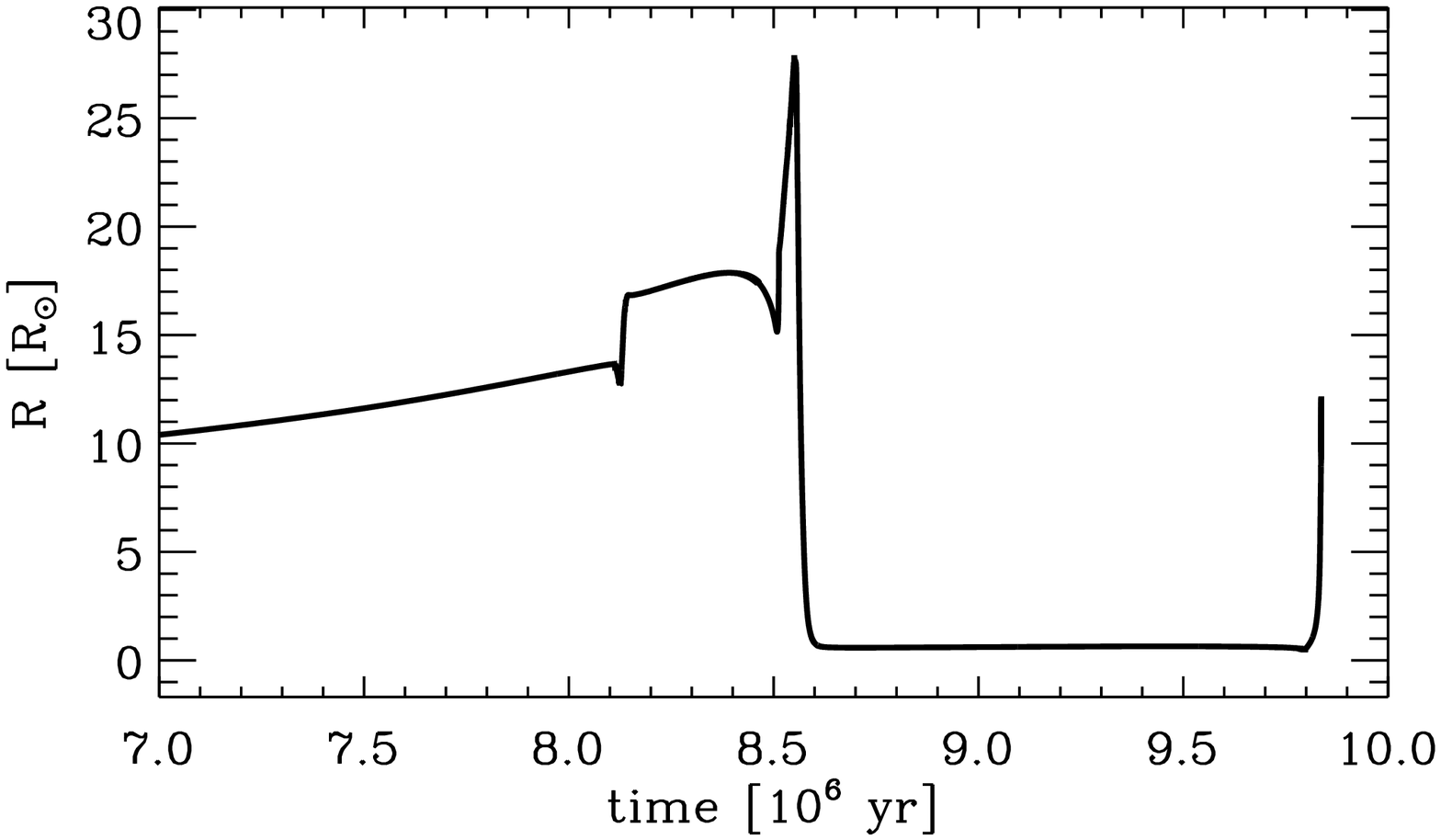}
\caption{\emph{Top panel}: 
Mean specific angular momentum of the innermost $3~\mathrm{M_\odot}$ and
$1.4~\mathrm{M_\odot}$ of the primary star in Seq.~14 as a function of the
evolutionary time. The time spans for the Case A and AB mass transfers are
marked by the color shades as indicated by the labels.  
\emph{Second panel}: Distribution of angular momentum in the primary star in
Seq.~9 at different evolutionary stages, as indicated by the labels.
\emph{Third panel}:
The orbital separation of the binary system in Seq.~14, as a function of time.
\emph{Bottom panel}:
The radius change of the primary star in Seq.~14, as a function of time.
}\label{fig:jcoreseq9} \end{figure}

\subsection{Fiducial Models}\label{sect:fiducial}
\subsubsection{Evolution with Case A and AB mass transfers}

The evolution of the primary star in a close binary system is characterized by
the rapid loss of mass due to Roche-lobe overflow.  As an example, the
evolution of the primary star in Seq.~14 is described in
Figs.~\ref{fig:kippseq9} and~\ref{fig:hrseq9},  where our fiducial value of
$f_\mathrm{sync}=1$ is adopted, including the Spruit-Tayler dyanamo.  The
binary system initially consists of a $18~\mathrm{M_\odot}$ star and a
$17~\mathrm{M_\odot}$ star in a 4 day orbit.  Mass transfer starts at $t= 8.09
\times 10^{6}~\mathrm{yr}$, when the helium mass fraction in the hydrogen
burning core has increased to 0.94. The mass transfer rate rises up to
$8\times10^{-4}~\mathrm{M_\odot yr^{-1}}$, which roughly corresponds to
$M_\mathrm{1}/\tau_\mathrm{KH,1}$ where $M_\mathrm{1}$ and $\tau_\mathrm{KH,1}$
denote the mass and the Kelvin-Helmoltz time scale of the primary star,
respectively. The primary mass decreases to $7.5~\mathrm{M_\odot}$ by the end
of the Case~A transfer (see Fig.~\ref{fig:chem}).  The second Roche-lobe
overflow begins at $t= 8.513\times 10^{6}~\mathrm{yr}$ when the envelope of the
primary star expands due to hydrogen shell burning during the helium core
contraction phase (Case~AB mass transfer). The primary star loses most of the
hydrogen envelope as a result,  exposing its helium core of
$3.95~\mathrm{M_\odot}$ having a small amount of hydrogen  ($M_\mathrm{H} =
0.04$~\Msun) in the outermost layers, as shown in the third panel of
Fig.~\ref{fig:chem}. 

Although the star remains compact ($R < 0.9~\mathrm{R_\odot}$) during core
helium burning, helium shell burning activated after core helium exhaustion
leads to the expansion of the envelope up to $\sim 12~\mathrm{R_\odot}$ (see
Fig.~\ref{fig:hrseq9}) during core carbon burning.  A Case~ABB mass transfer
does not occur, however, due to the large orbital separation ($A = \sim
121~\mathrm{R_\odot}$) at this stage, while it does occur in many other
sequences.  The final mass at the end of the calculation (neon burning) is
$3.79~\mathrm{M_\odot}$.  The mass of hydrogen decreases to  0.0015~\Msun{} at
the end, and the remaning mass of helium is 1.49~\Msun, as shown in the last
panel of Fig.~\ref{fig:chem}.  The star is likely to eventually explode as a
Type Ib supernova given the rather thick helium envelope with  a very thin
hydrogen layer, but it might also appear as Type IIb if the supernova were
found within several days after the explosion (see Sect.~\ref{sect:hydrogen}).

\begin{figure}
\epsscale{1.0}
\plotone{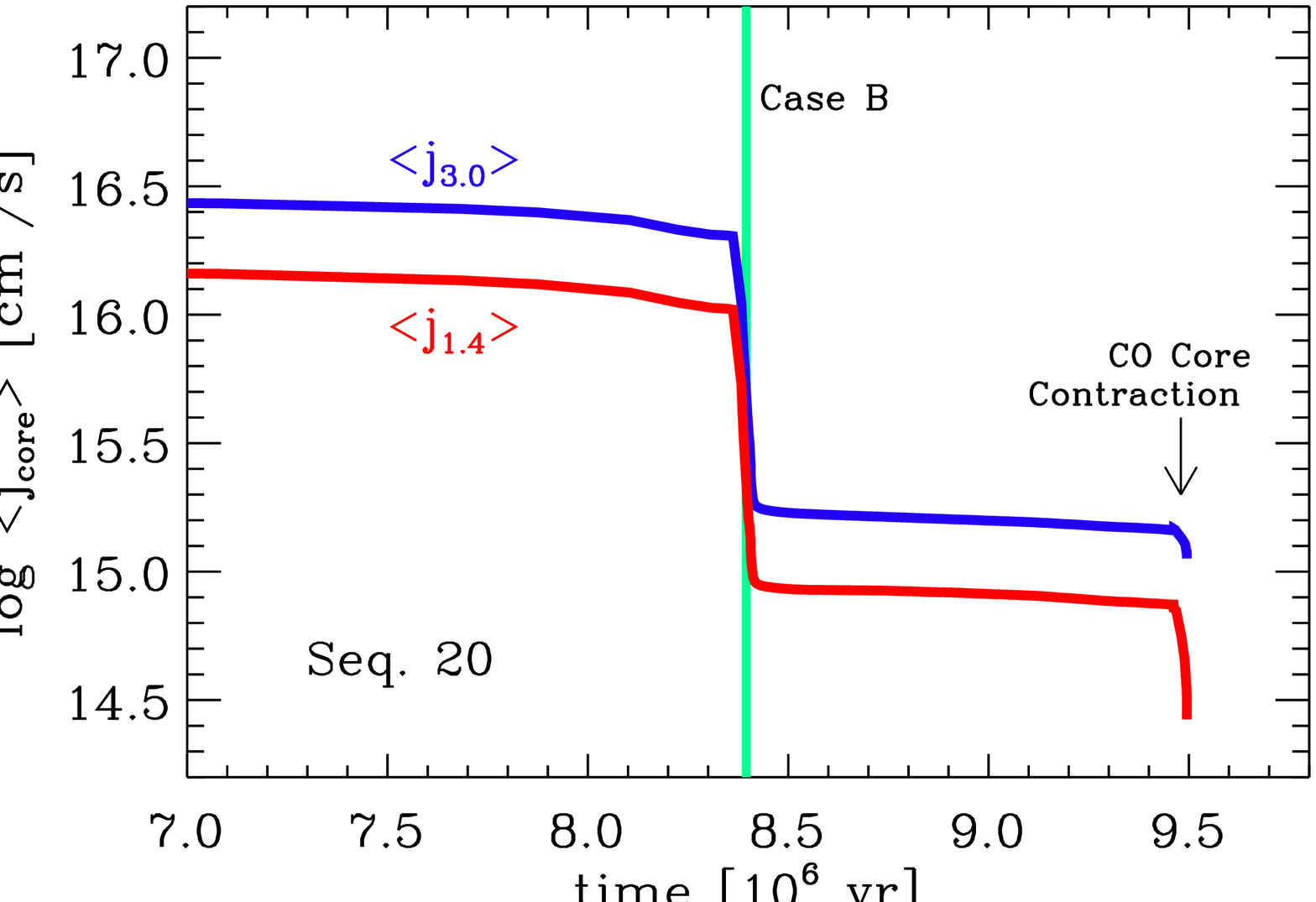}
\caption{Mean specific angular momentum of the innermost $3~\mathrm{M_\odot}$ and $1.4~\mathrm{M_\odot}$
of the primary star in  Seq.~20, 
as a function of the evolutionary time. The time spans for the Case B mass transfers are marked by 
the color shades as indicated by the labels.
}\label{fig:caseb}
\end{figure}

\begin{figure}
\epsscale{1.0}
\plotone{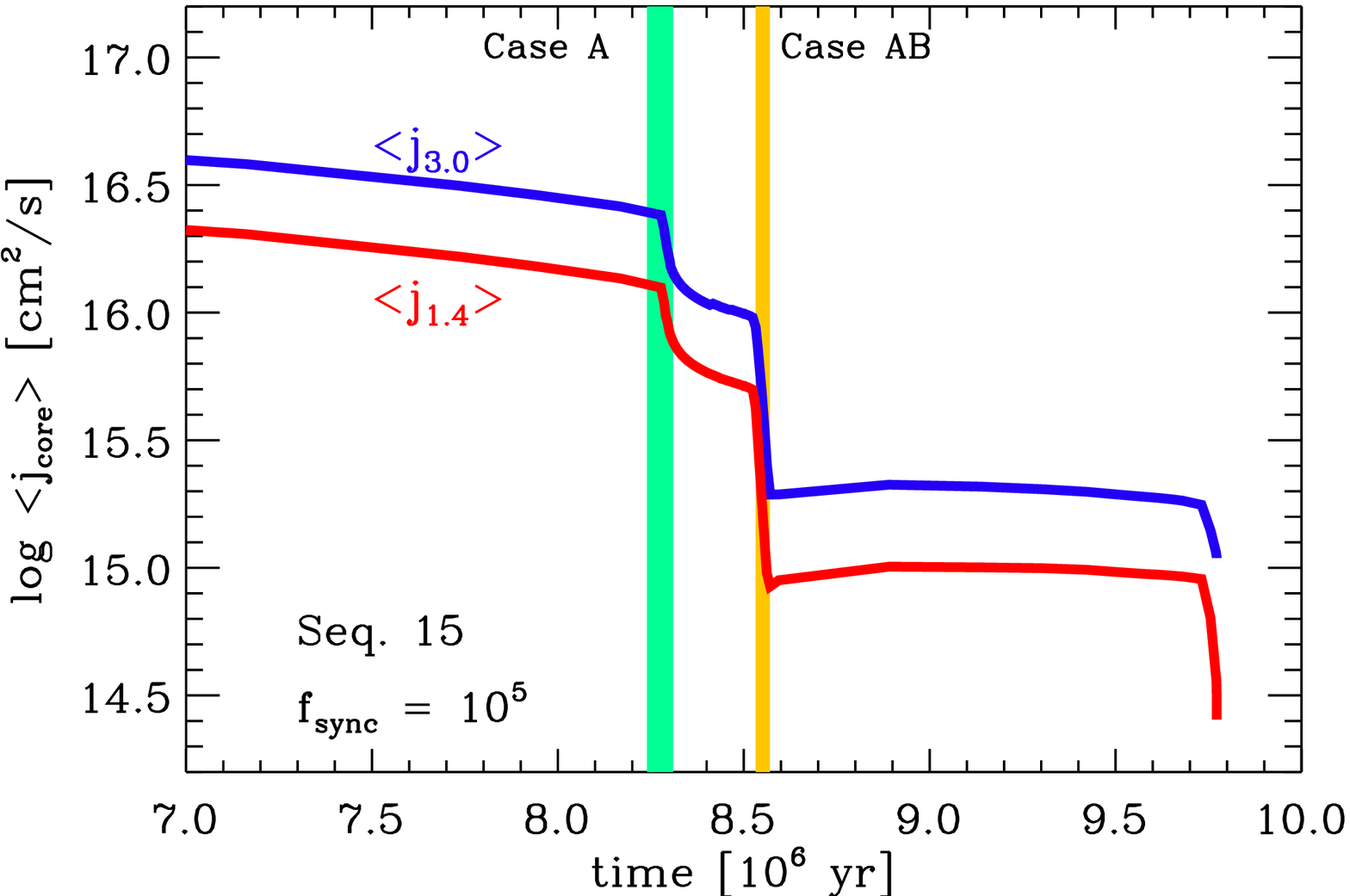}
\plotone{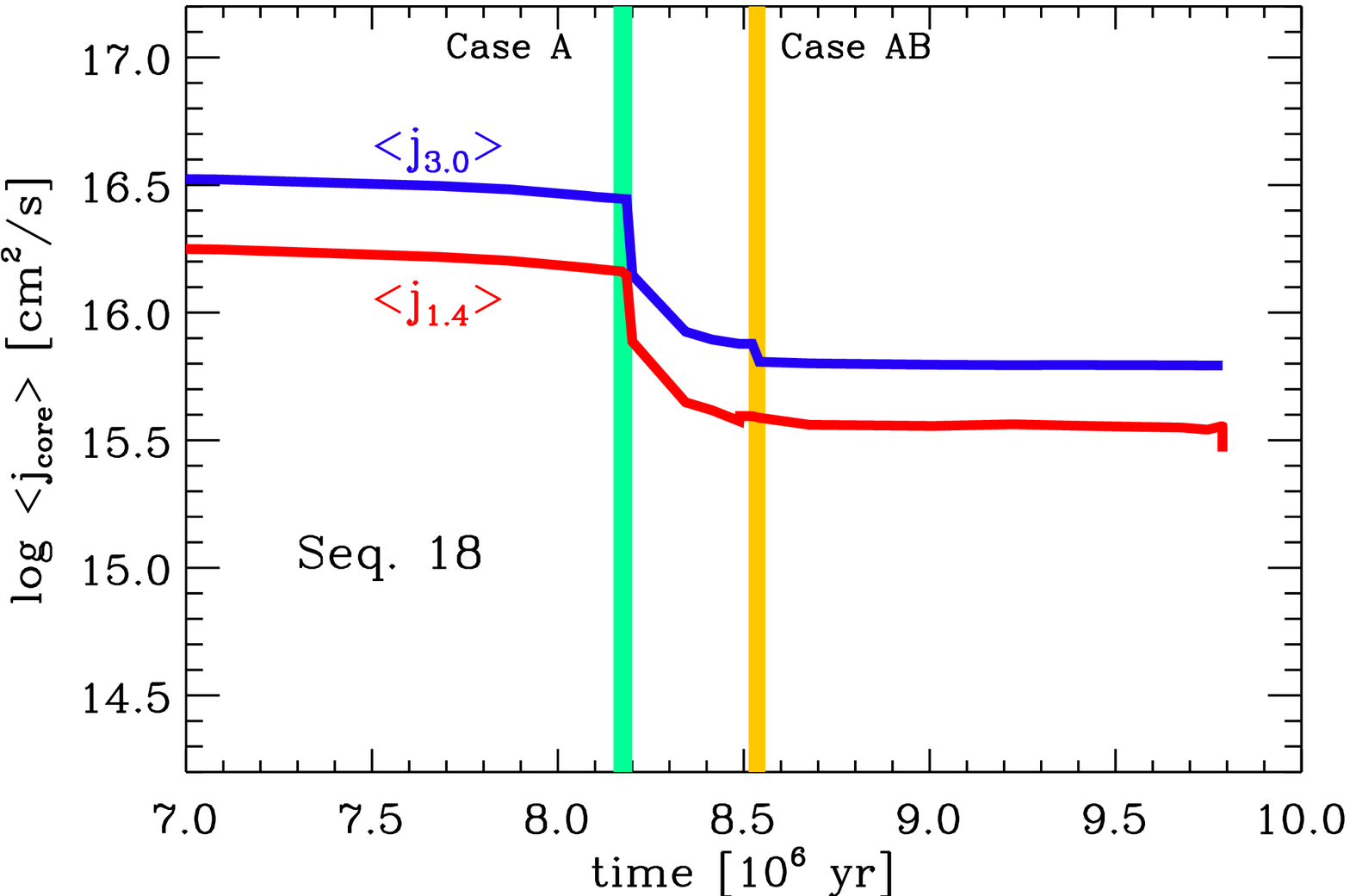}
\caption{Mean specific angular momentum of the innermost $3~\mathrm{M_\odot}$ and $1.4~\mathrm{M_\odot}$
of the primary star in Seq.~15 (top) and Seq.~18 (bottom), 
as a function of the evolutionary time. The time spans for the Case A and AB mass transfers are marked by 
the color shades as indicated by the labels.
}\label{fig:jcoreseq10}
\end{figure}

The lower panel of Fig.~\ref{fig:hrseq9} shows that the mass transfer is not
conservative.  When the secondary star reaches critical rotation as a
result of the accretion of angular momentum, the stellar wind mass
loss rate increases so drastically as to prevent efficient mass accumulation
\citep[see][for a more detailed discussion on this effect]{Petrovic05b}.  The
ratio of the accreted mass in the secondary star to the transferred mass
from the primary star is about 0.83 during the Case~A mass transfer, and 0.41
during the Case~B mass transfer.  
The previous calculations by WL99 and \citet{Wellstein01b} show that the
non-conservative mass transfer leads to a shorter orbit than in the case of
conservative mass transfer, in general.  This effect should be kept in mind
in the following discussion on the evolution of the binary orbit and its
consequences. 

The evolution of rotation in the primary star is shown in
Figs.~\ref{fig:totjsync}, \ref{fig:omegasync} and \ref{fig:jcoreseq9}.  As
shown in Fig.~\ref{fig:totjsync}, the total angular momentum of the primary
star rapidly decreases in the beginning as a result of tidal interaction, until
the star is completely synchronized with the orbit when $t\simeq
2\times10^5~\mathrm{yr}$.  Fig.~\ref{fig:omegasync} shows the evolution of the
angular velocity inside the primary star from ZAMS until it reaches complete
synchronization. Note that the star keeps rotating almost rigidly,  even though
the tidal interaction causes redistribution of angular momentum from the
outermost layers of the star~\citep{Wellstein01, Detmers08}.  This results from the very
short time scale of the transport of angular momentum in the star (about 300 years) 
due mainly to convection and the Spruit-Tayler dynamo, in the convective and radiative
layers respectively. 

Later on, the angular velocity of the primary star still remains coupled with
the orbital motion, until decoupling starts during Case B mass transfer (see
below). Interestingly the total angular momentum of the primary star gradually
increases from the initial synchronization until the onset of Case A mass transfer
(Fig.~\ref{fig:totjsync}).  This is because the star significantly
expands, while the change of the orbital separation remains small durning this
period, as shown in the third and last panels in Fig.~\ref{fig:jcoreseq9}. 
However, it rapidly decreases again during the Case A and AB mass transfer
phases, as explained below.

Fig.~\ref{fig:jcoreseq9} shows that angular momentum in the core of the primary
star is mostly removed during the mass transfer phases (Case~A and Case~AB).
It should be noted, however, that the mechanism for braking the core 
is different for each case. During the Case A mass transfer, the synchronization time
scale according to Eq.~(\ref{eq3}) remains very short ($~10^3~\mathrm{yr}$)
compared to the mass transfer time scale ($~10^4~\mathrm{yr}$).  The decrease
of the core angular momentum during the Case A mass transfer results from the
synchronization that occurs even when the orbit is rapidly widened due to mass
exchange (Fig.~\ref{fig:jcoreseq9}).  During the Case AB mass transfer, the further
increase of the orbital separation significantly weakens the role of
synchronization for the redistribution of angular momentum.  However, both the
Spruit-Tayler dynamo and mass loss lead to rapid braking of the thermally
contracting helium core. Further significant core-braking by the Spruit-Tayler
dynamo occurs during the CO core contraction phase, and the mean specific
angular momentum of the innermost $1.4~\mathrm{M_\odot}$ becomes about
$2.5\times10^{14}~\mathrm{cm^2~s^{-1}}$ at the neon burning phase.  

\subsubsection{Evolution with Case B mass transfer}

In Seq.~20, the initial masses of the stellar components are the same as in
Seq.~14 but the initial period is large enough that the first mass transfer
occurs during the helium core contraction phase (Case B mass transfer).
Avoiding Case A mass transfer,  the primary star retains more angular momentum
in the core at the end of main sequence than in Seq.~14. However, the core loses
more angular momentum during the helium core contraction phase than in Seq.~14,
as stronger magnetic torques are exerted, mainly due to the more massive
envelope (Fig.~\ref{fig:caseb}).  At neon burning, $<j_\mathrm{1.4}> = 2.6
\times 10^{14}~\mathrm{cm^2~s^{-1}}$ is obtained, which is similar to that in
Seq.~14.

\subsection{Non-fiducial Models}\label{sect:nonfiducial}
\subsubsection{Influence of the synchronization time scale}

The effect of synchronization is negligible in Seq.~15, where $f_\mathrm{sync}
= 10^5$ is adopted.  As shown in Fig.~\ref{fig:jcoreseq10}, a rather rapid
decrease of $<j_\mathrm{core}>$ occurs during and after the Case A mass
transfer, which is a combined effect of mass loss and magnetic torques: mass
loss carries away angular momentum from the envelope, and magnetic torques
brake the core rotation subsequently.  The core is further slowed down during the
helium core and CO core contraction phases, resulting in $<j_\mathrm{1.4}> =
2.6 \times 10^{14}~\mathrm{cm^2~s^{-1}}$ at the neon burning phase.  Note that
this value is very close to that in Seq.~14. 

We find that the result with Zahn's prescription for synchronization (Seq. 17)
is not much different from that of Seq.~15 where synchronization is
negligible.  The synchronization time scale according to Zahn is sensitive to
the ratio of the convective core size to the stellar radius
($1/\tau_\mathrm{sync} \propto (R_\mathrm{conv}/R_\mathrm{star})^8$, see
Eqs.~(\ref{eq4}) and (\ref{eq5})).  This ratio continuously decreases as the
star evolves, and thus $\tau_\mathrm{sync}$ continuously increases to such an
extent that the effect of synchronization can be ignored when Case A mass
transfer starts.  When $f_\mathrm{sync} = 0.01$ with Tassoul's prescription is
used (Seq. 16), on the other hand, synchronization becomes important even after
the Case AB mass transfer phase, and the primary is further spun down by tidal
interaction, giving $<j_\mathrm{1.4}> = 6~10^{13} \times \mathrm{cm^2~s^{-1}}$
at the neon burning phase. 

\subsubsection{Non-magnetic model}

In Seq.~18 where the Spruit-Tayler dynamo is not included, the core is spun
down due to synchronization during the Case A mass transfer phase
(Fig.~\ref{fig:jcoreseq10}) as in the corresponding magnetic case (Seq.~14;
Fig.~\ref{fig:jcoreseq9}).  The spin-orbit coupling becomes significantly
weakened as the orbit widens after the Case A mass transfer phase.  Despite
a significant amount of mass is lost via Case AB mass transfer, the core in the
primary star retains most of the remaining angular momentum in the following
evolutionary stages.  This is because the chemical gradient across the boundary
between the helium core and the hydrogen envelope effectively prohibits
the transport of angular momentum \citep[cf.][]{Meynet98, Heger00a}.  The core
angular momentum at neon burning is thus about 10 times larger
($<j_\mathrm{1.4}> = 3.57~10^{15}~\mathrm{cm^2~s^{-1}}$) than in the
corresponding magnetic case.

\subsection{Discussion}

As shown in the above examples, in the model sequences with the Spruit-Tayler
dynamo, all of the primary stars retain similar amounts of angular momentum (a
few $10^{14}~\mathrm{cm^2~s^{-1}}$) in the innermost $1.4~\mathrm{M_\odot}$ at
neon burning regardless of the detailed history of mass transfer, unless
synchronization is extremely fast as in Seq.~16 (see Table~\ref{tab1}).
According to the Spruit-Tayler dynamo, magnetic torques exerted to the core
become stronger with a higher spin rate, a larger degree of differential
rotation between the core and the envelope, and a heavier radiative envelope.
Therefore, although winds or Roche-lobe overflows reduce the size of the
hydrogen envelope and remove angular momentum from the star, this in turn
weakens the torque exerted to the core, and vice versa.  The remarkable
convergence of   $<j_\mathrm{1.4}>$ to a few $10^{14}~\mathrm{cm^2~s^{-1}}$ in
our model sequences, even for different wind parameters and metallicities as shown in Table~\ref{tab1}, 
can be explained by this self-regulating nature of the
Spruit-Tayler dynamo.  This result indicates that not much diversity is
expected in SNe Ibc progenitors produced via Case~A or Case AB/B mass
transfer, in terms of rotation:  most of SNe Ibc of a similar progenitor mass
may leave neutron stars with a similar spin rate.  However, other types of
binary interactions still may lead to various final rotation periods in SN
progenitors \citep[see][for such examples]{Brown00, Cantiello07, Podsiadlowski10}.

\section{The nature of SN Ibc progenitors}\label{sect:sn}

\begin{figure}
\epsscale{1.}
\plotone{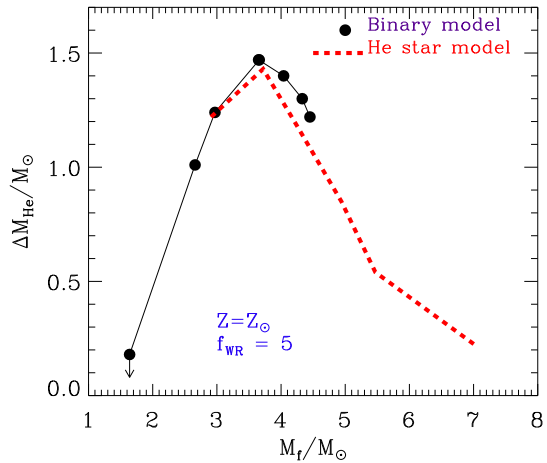}
\caption{
 The amount of helium in SN Ibc progenitor models at $Z=\mathrm{Z_\odot}$ with
$f_\mathrm{WR} =5$, as
function of the final mass.  The filled circles denote the
prediction from our binary star models.  
The results from mass-losing pure helium
star models are marked by dashed line.
}\label{fig:dmhe} 
\end{figure}

In the literature, the detailed history of binary interactions such as mass
transfer and tidal interaction is often neglected, and only the evolution of
pure helium stars to discuss SNe Ibc progenitors in binary systems is
investigated (e.g. WLW95; \citealt{Pols02}).  Although this approach is very
useful, our binary star models indicate that complications resulting from
binary interactions have important consequences in SN Ibc progenitors.  To
address this issue, here we also present evolutionary models of non-rotating
single helium stars  with initial masses of 2.8 -- 20~\Msun{} at $Z=0.02$  for
comparison with binary star models, as presented in Table~\ref{tab2}.  This
also enables us to study SN Ibc progenitors with initial masses higher than
25~\Msun, which are lacking in our binary model sequences.  These helium star
models were calculated up to neon burning, with the WR mass loss rate given in
Eq.~(\ref{eq1}) with $f_\mathrm{WR} =$ 5 or 10, as in the binary models. 
In the following, we focus our discussion mostly on the results with $f_\mathrm{WR} = 5$,
while the influence of the wind parameter is briefly discussed.

\subsection{Final mass}\label{sect:masses}

The final mass of the primary stars  in a close binary systems is largely
determined by the winds mass loss during core helium burning,  if the initial
mass ($M_\mathrm{He,i}$; i.e., the mass right after Case AB or Case B mass
transfer) is significantly larger than about 3.0~\Msun{}.  However, for a less
massive helium star, the expansion of the helium envelope becomes so dramatic
during CO core contraction, and/or during core carbon burning, that it can lead
to another mass transfer phase: Case ABB or Case BB (e.g., compare the final
radius of the 2.8 \Msun{} helium star model with those of more massive helium
star models in Table~\ref{tab2}).

The impact of this mass transfer phase becomes more important for a less
massive helium star. In Seqs. 1, 2 and 4 where $M_\mathrm{He,i} \simeq$ 2.1 --
2.3~\Msun{},  the rapid loss of mass during Case ABB/BB transfer reduces the
total mass of the primary stars to such an extent that they may not explode as
supernovae, but die as white dwarfs. In Seq.~3, where $M_\mathrm{He,i} \simeq
2.7$~\Msun{}, the carbon-oxygen core can grow beyond the Chandrasekhar limit
despite the significant loss of mass (i.e., about 1~\Msun) during the Case BB
phase.  The remaining helium mass in the envelop is only about 0.18~\Msun{} at core
carbon exhaustion. For the other sequences at solar metallicity (Seqs. 5, 8, 9,
10, 11, 13, 21) where $M_\mathrm{He, i} > 3.0~\mathrm{M_\odot}$,  the amounts of
mass loss via Case ABB/BB transfer until core neon burning are rather moderate,
varying from 0.02 to 0.1~\Msun{}.  

The WR wind mass loss rate has significant consequences in 
systems where  $M_\mathrm{He,i} > $ 3.0~\Msun{}.  The previous study by  WLW95
concluded that the final masses of single helium stars with initial masses of 3
-- 20~\Msun{} should converge to $\sim$ 2.2 -- 3.5~\Msun{}.  As expected from
the lower WR mass loss rate adopted in this study, our helium
star models give a significantly wider range of final masses ($M_\mathrm{f}$)
at the given metallicity of 0.02: $M_\mathrm{f}=$ 2.9 -- 7~\Msun{} (2.9 --
10~\Msun) from helium stars of 3 -- 20~\Msun{} for $f_\mathrm{WR} = 5$
($f_\mathrm{WR}=10$; Table~\ref{tab2}; cf.  Fig.~\ref{fig:mimf}).  Our binary
star models also give generally larger final masses for SN Ibc progenitors
than those in WL99: $M_\mathrm{f} \simeq$ 1.64 -- 4.5~\Msun{} for $f_\mathrm{WR} =5$ from
$M_\mathrm{ZAMS} =$13 -- 25~\Msun{}, compared to  $M_\mathrm{f} \simeq$ 2.0 --
3.4~\Msun{} in WL99\footnote{Here, the lower end of $M_\mathrm{f}$ is
determined by the detailed history of Case ABB/BB transfer, which depends on
uncertain parameters such as the semi-convection efficiency.}.

\subsection{Helium}\label{sect:helium}

Figure~\ref{fig:dmhe} shows the amount of helium ($M_\mathrm{He}$) as a
function of the final mass in the SN Ibc progenitor models with $f_\mathrm{WR}
= 5$ at solar metallicity.  In general, both binary and helium star models
predict large amounts of helium in the envelope compared to the results of
WLW95 and WL99 except for relatively low mass systems of $M_\mathrm{He, i} \la
3.0$~\Msun{}.  Note that only rather massive progenitors with $M_\mathrm{f} \ga
5.5$~\Msun{} can be significantly helium deficient (i.e., $M_\mathrm{He} <
0.5~\mathrm{M_\odot}$). 

For the systems with  $M_\mathrm{f} \la 3.0$~\Msun{}, the role of Case ABB/BB
mass transfer becomes important. 
In particular, the
small amount of helium ($M_\mathrm{He} \la  0.18~\mathrm{M_\odot}$) in Seq. 3
shows that helium deficiency can also be achieved by the so-called Case BB/ABB
mass transfer from relatively low mass helium cores ($M_\mathrm{He,i} <
3~\mathrm{M_\odot}$). Therefore, we expect two different classes for Type Ic
progenitors at solar metallicity: one with $M_\mathrm{f} \la 2.0$~\Msun{} and
the other with $M_\mathrm{f} \ga$ 5.5~\Msun{} \citep[see also][]{Wellstein99, Pols02}. 

Obviously, the upper limit of $M_\mathrm{f}$ for potential Type Ic progenitors increases
with a smaller mass loss rate:  with $f_\mathrm{WR} = 10$, we have
$M_\mathrm{f} \ga$ 8.9~\Msun{}.

\subsection{Hydrogen}\label{sect:hydrogen}

\begin{figure}
\epsscale{1.}
\plotone{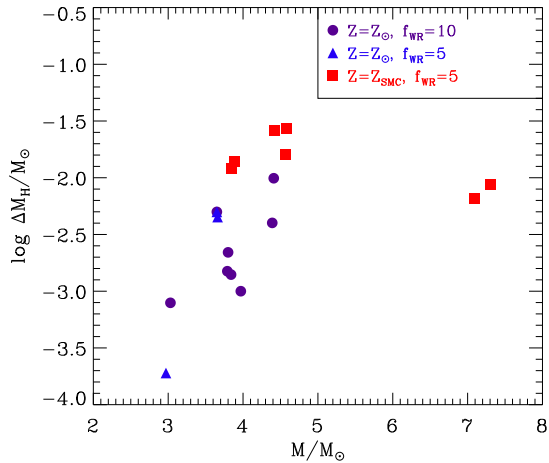}
\caption{
The amount of hydrogen in SN Ibc progenitor models as a function of the final
mass.  The filled triangles and circles denote the results of our binary star
models at $Z=$\Zsun{} with $f_\mathrm{WR}=$  5 and 10,  respectively. 
The fille squares give the results with the SMC metallicity models.}\label{fig:dmh} 
\end{figure}

In our binary star models, Case AB or Case B mass transfer does not completely
remove hydrogen from the primary stars, as shown in Fig.~\ref{fig:chem}.  The
resulting WR mass loss is therefore weaker than that from the corresponding
pure helium star.  This explains the fact that for a given WR mass loss rate,
the binary star models with $M_\mathrm{f} \ga 3.0 $~\Msun{} predict somewhat
larger  $M_\mathrm{He}$ than the single helium star models do as shown in
Fig.~\ref{fig:dmhe}. 

 More importantly, the binary star models show that, even at $Z\approx
Z_\mathrm{\odot}$, small amounts of hydrogen can be retained up to the
pre-supernova stage for a certain range of the final mass.  Fig.~\ref{fig:dmh}
(see also Table~\ref{tab1} and Fig.~\ref{fig:chem}) shows the total mass of
hydrogen in the primary stars at neon/oxygen burning phase.  The time span from
this stage to core collapse is supposed to be less than about 10 yr.  The WR
winds mass loss rate from the primary star models shown in the figure is about
$10^{-6}~\mathrm{M_\odot~yr^{-1}}$.  Some of the primary stars in this sample
are still undergoing Case ABB or BB mass transfer (Seqs.  5, 8, 9, 13, \& 21
and many of the SMC models; see Table~\ref{tab1}), but the mass transfer rate
is only about $10^{-5}~\mathrm{M_\odot~yr^{-1}}$.  Therefore, the amount of
hydrogen at core collapse should remain close to the values given in the
figure.

Fig~\ref{fig:dmh} indicates that the presence of hydrogen is only expected for
$ 3.0 \la M_\mathrm{f} [\mathrm{M_\odot}] \la 3.7$, at solar metallicity with
$f_\mathrm{WR}=5$.  This results from two different reasons for different mass
ranges. For $M_\mathrm{f} < 3.0~\mathrm{M_\odot}$, primary stars expand during
the carbon burning phase to much larger radii than more massive ones do.  The
resulting  mass loss rates via Case ABB or BB transfer thus become large enough
to completely remove hydrogen from the primary stars by the time of core
collapse.  For $M_\mathrm{f} \ga 3.7~\mathrm{M_\odot} $, on the other hand, WR
winds are rather strong and can remove hydrogen from the primary stars during
the core helium burning phase.  

The upper limit of  $M_\mathrm{f}$ for the presence of hydrogen thus depends on
the adopted mass loss rate. As shown in the figure, with $f_\mathrm{WR} = 10$, it
increases to about 4.5~\Msun{} at solar metallicity. At SMC metallicity, the
stellar wind effect is less significant and even rather massive supernova
progenitors of about 7~\Msun{} can retain fairly large amounts of hydrogen
($\sim 10^{-2}~\mathrm{M_\odot}$). We discuss implications of this result 
for supernova types in Sect.~\ref{sect:dischydrogen}.

\subsection{Radius}

\begin{figure}
\epsscale{1.}
\plotone{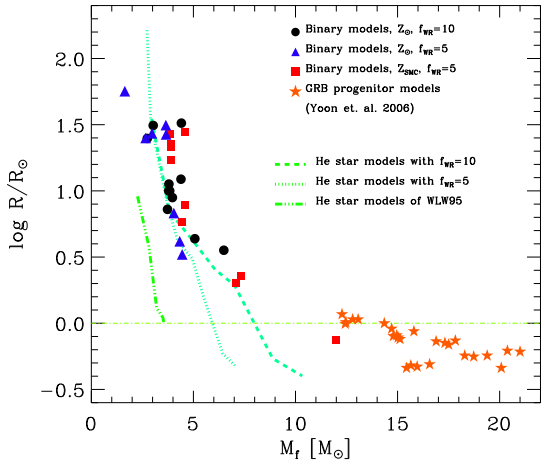}
\caption{The predicted radii of SN Ibc progenitors as a function of the final
mass. The filled circles and triangles denote the results of our binary star models at
$Z=\mathrm{Z_\odot}$ for $f_\mathrm{WR} =$ 10 and 5, respectively, while the filled squares are
for $Z=\mathrm{Z_{SMC}}$.  
The results from mass-losing single helium star models at $Z=\mathrm{Z_\odot}$ are marked
by the dashed ($f_\mathrm{WR} = 10$), dotted ($f_\mathrm{WR} = 5$) and dashed-three-dotted (WLW95) lines. 
The thin dashed-dotted line gives the solar radius. 
}\label{fig:radius} 
\end{figure}

Relatively low mass helium stars usually experience a rapid expansion of
the envelope during the core carbon burning phase.  Interestingly, our new
models predict much larger radii at the pre-supernova stages than what the WLW95 models do,
as shown in Fig.~\ref{fig:radius}. The first reason for this difference is the
updated OPAL opacity table by \citet{Iglesias96}, which causes a strong iron
bump at around $\log T = 5.3$.  
Another reason is the smaller WR mass loss rate adopted in the present study, and
the resultant thicker helium envelope in our models.  The radius is further affected by the
complication of binary interaction: the presence of a thin hydrogen layer in
some of the binary models results in a more extended envelope than in the
corresponding single helium star models.  It is also worth noting that the
radii of the binary star models at $M_\mathrm{f} < 3~\mathrm{M_\odot}$ are
smaller than those of the single helium star models.  This is because the
primary stars in these sequences are filling the Roche-lobe, and the envelope
cannot expand beyond it.  

The metallicity effect is somewhat subtle.  The hydrogen and helium layers in
SN Ibc progenitors become thicker due to the reduced mass loss rate for lower
metallicity, while the opacity due to metals becomes smaller. As these two
effects compensate each other, the radii of our SN Ibc progenitor
models at both SMC and solar metallicities are found to be similar.  Finally,
it should be noted that the radius depends on the final mass. In general, a
less massive progenitor tends to have a larger radius at the presupernova
stage.  This might make mixing of nickel into the helium rich layer induced by the
Rayleigh-Taylor instability during the supernova explosion more
efficient for a less massive SN Ibc progenitor, as implied by the simulations
of \citet{Hachisu91} and \citet{Joggerst09}\footnote{However, it should be kept
in mind that the degree of mixing due to the Rayleigh-Tayler instability may
depend not only on the stellar structure, but also on explosion energy and
directional asymmetry \citep{Hammer09}}, resulting in more prominent helium
lines in the spectra (see Sect.~\ref{sect:discussion} below).

\subsection{Winds}

\begin{figure}
\epsscale{1.}
\plotone{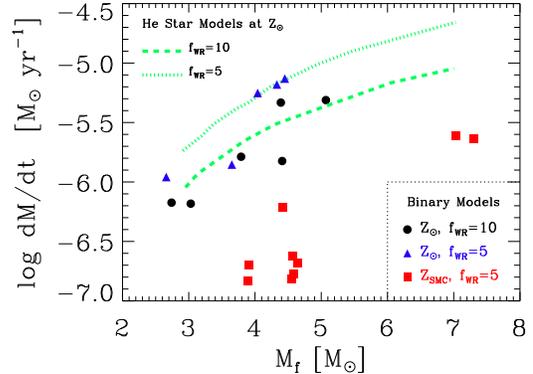}
\caption{The predicted winds mass loss rates of SN Ibc progenitors as a function of the final
mass. The filled circles and triangles denote the results of our binary star models at
$Z=\mathrm{Z_\odot}$ for $f_\mathrm{WR} =$ 10 and 5, respectively, while the filled squares are
for $Z=\mathrm{Z_{SMC}}$.  
The results from mass-losing single helium star models at $Z=\mathrm{Z_\odot}$ are marked
by dashed ($f_\mathrm{WR} = 10$) and dotted ($f_\mathrm{WR} = 5$) lines. 
}\label{fig:mdot} 
\end{figure}

Fig.~\ref{fig:mdot} shows the mass loss rate due to WR winds at the central
neon/oxygen burning phase from different SN Ibc progenitor models.  The mass
loss rate ranges from $10^{-6}~\mathrm{M_\odot yr^{-1}}$ to $\sim
10^{-5}~\mathrm{M_\odot yr^{-1}}$, depending on the final mass,  at solar
metallicity.  The corresponding wind velocity (i.e.,
the escape velocity) changes from $\sim 200~\mathrm{km~s^{-1}}$ to $\sim
2400~\mathrm{km~s^{-1}}$. 

The circumstellar interaction of such winds from primary stars in binary
systems are supposed to be very complex because of the orbital motion and the
wind-wind interaction with the secondary star.  Some of the primary star models
shown in the figure are still undergoing Case ABB or BB mass transfer, which
may add to the complexity.  Therefore, the nature of the circumstellar
materials around SN Ibc progenitors in a binary system may not follow the
simple relation of $\rho \propto r^{-2}$ that is expected for single WR
star progenitors. 

\section{Discussion}\label{sect:discussion}

We have presented new evolutionary models of massive close binary stars,
considering tidal interaction, and transport of angular momentum and chemical
species due to rotationally induced hydrodynamic instabilities and the
Spruit-Tayler dynamo. We have investigated the redistribution of angular
momentum in the primary star. Although mass transfer and tidal interaction can
significantly affect the evolution of the rotation velocity of the primary star
on the main sequence, the amount of angular momentum retained in the core in
the late evolutionary stages is rather insensitive to the previous history of
such binary interactions because of the self-regulating nature of the
Spruit-Tayler dynamo. 

We have also calculated non-rotating, mass-losing single helium star models and compared them with
our primary star models in binary systems.  Our models adopt a much lower WR mass loss rate than in
the previous studies by  WLW95 and WL99, and predict some new important properties of SN Ibc
progenitors accordingly. 
The following discussions are based on the models with $f_\mathrm{WR} = 5$, unless
otherwise specified.

\begin{enumerate}

\item  The final masses of SN Ibc progenitors in binary systems at
$Z\simeq$~\Zsun{} are not limited to $1.5~\mathrm{M_\odot} \la M_\mathrm{f} \la
4~\mathrm{M_\odot}$ as predicted by WLW95 and WL99, but a more wide range 
$M_\mathrm{f}$ is expected (i.e., $1.5~\mathrm{M_\odot} \la M _\mathrm{f} \la
7.1~\mathrm{M_\odot}$ from $M_\mathrm{init} \simeq 12 ... 60$~\Msun; see Fig.~\ref{fig:mimf}).

\item At $Z \simeq$~\Zsun, significant deficiency of helium ($M_\mathrm{He} <
0.5$~\Msun{}) is expected for $M_\mathrm{f}/\mathrm{M_\odot} \ga 5.5$ and for
$1.5 \la M_\mathrm{f}/\mathrm{M_\odot} \la 2.0$ (Fig.~\ref{fig:dmhe}).  Rather
large amounts of helium up to 1.5~\Msun{} are expected for the other final mass
range (i.e., $2.0 \la M_\mathrm{f}/\mathrm{M_\odot} \la$ 5.5~\Msun{}). 
At $Z \simeq$~\Zsmc, no such helium deficient SN progenitors are expected
for the considered initial masses ($16 - 40$~\Msun). 

\item A thin layer of hydrogen with $M_\mathrm{H} = 10^{-4} - 
10^{-2}$~\Msun{} is predicted for SN Ibc progenitors with  $
3.0~\mathrm{M_\odot} \la M_\mathrm{f} \la 3.7~\mathrm{M_\odot}$ at $Z \simeq
$\Zsun, and  $ 3.0~\mathrm{M_\odot} \la M_\mathrm{f} \la 8 ~\mathrm{M_\odot}$
at $Z \simeq$\Zsmc, respectively (Fig.~\ref{fig:dmh}; Table~\ref{tab1}). 

\item Most SN Ibc progenitors with $M_\mathrm{f} \la 5 $~\Msun{} rapidly
expand during core carbon burning, resulting in  $ R =  \sim 4.0 -
30~\mathrm{R_\odot}$ at the presuprnova stage.  This is much larger than found
in WLW95 (Fig.~\ref{fig:radius}). Compact progenitors of $R \la
~\mathrm{R_\odot}$ are only expected for a relatively high mass ($M_\mathrm{f}
\ga 5.5$~\Msun{} at $Z\simeq$~\Zsun{} and $M_\mathrm{f}
\ga 10$~\Msun{} at $Z\simeq$~\Zsmc{} 
; Fig.~\ref{fig:radius}). 
\end{enumerate}

The above results raise several important issues regarding observational
consequences, as discussed below. 

\subsection{Implications for energetic explosions powered by rapid rotation}

Our binary star models show that the mass transfer during helium core
contraction (Case AB or Case B) in a close massive binary system cannot remove
the hydrogen envelope promptly enough to avoid the core braking due to the
Spruit-Tayler dynamo during the helium core contraction phase.  
Comparison of our binary star models with the single
star models by \citet{Heger05} and \citet{Yoon06} indicate that the amount of
angular momentum retained in the core of the primary star at the presupernova
stage should not be much different from those found in single star models if
the Spruit-Tayler dynamo is adopted. I.e., a specific angular momentum of a few
$10^{14}~\mathrm{cm^2~s^{-1}}$ in the innermost $\sim$1.4~\Msun{} at the
presupernova stage is expected in both single and binary stars.  This value is
smaller by one or two orders of magnitude than what is necessary to make a long
gamma-ray bursts by magnetar or collapsar formation, or very energetic
supernovae (hypernovae) powered by rapid rotation and strong magnetic fields
~\cite[e.g.][]{Burrows07}, although it may suffice to produce millisecond
pulsars \citep{Heger05}. Together with the work by \citet{Petrovic05a},  our
results thus indicate that binary interactions with Case AB/B mass transfers at
$Z\approx \mathrm{Z_\odot}$ may not particularly enhance the production of
strongly rotation-powered events like long GRBs or hypernovae. This is
consistent with the observational evidence that such events are rare compared
to normal core collapse events \citep[e.g.,][]{Podsiadlowski04, Guetta07}.
This also confirms the theoretical consensus that other evolutionary
paths are needed to produce long GRBs associated with SN Ibc, such as the quasi-chemically homogeneous
evolution of a metal poor star \citep{Yoon05, Yoon06, Woosley06, Cantiello07}, 
tidal spin-up of a WR star in a very close binary system with a neutron star
or black hole companion \citep[e.g.][]{Izzard04, Heuvel07}\footnote{A
recent study using detailed stellar evolution models by \citet{Detmers08},
however, seriously questions this possibility.}, or binary evolution with Case C mass transfer
with some specific conditions \citep{Brown00, Podsiadlowski10}. 

\subsection{Progenitor size}

Larger radii of our SN Ibc progenitor models than those previously found
should have consequences in shock break-outs and bolometric light curves.  For
instance, a shock break-out from a larger envelope would be marked by a lower
photosperic temperature . Detailed comparison of numerical calculations with
observational data may thus give strong constraints on SNe Ibc progenitor
properties \citep[e.g.,][]{Calzavara04}.  Recent discovery of the X-ray outburst
with SN 2008D by \citet{Soderberg08} indeed suggests the usefulness of such a study
for the probe of supernova progenitors \citep[e.g.][]{Soderberg08, Chevalier08,
Xu08, Modjaz08}, for which our new models would provide ideal input.  We will
address this issue in a forthcoming paper.

\subsection{Presence of helium and implications for SNe Ic}\label{sect:snic}

\begin{figure}
\epsscale{1.}
\plotone{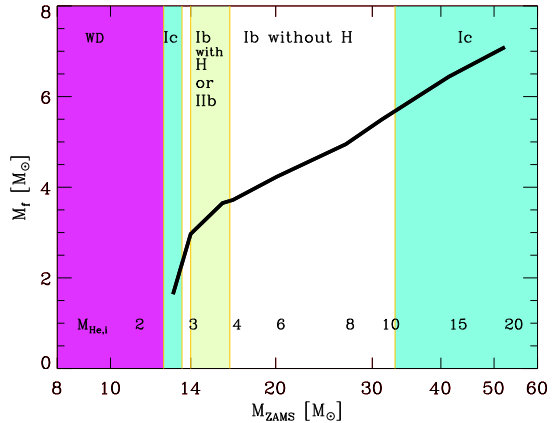}
\caption{
The predicted final masses of the primary stars in massive close binaries that
undergo Case B mass transfer, as a function of the zero-age main sequence
(ZAMS) mass, based on our binary and helium star models with $f_\mathrm{WR} =
5$.  The absicssa is given in log scale.  The numbers right above the abscissa
denote the initial masses of the helium stars that the primary stars of the
corresponding ZAMS masses would produce.  The expected final outcomes of the
primary stars according to different ZAMS masses are given by the labels right
below the top: white dwarf (WD), type Ic supernova (Ic), type Ib supernova
with a thin hydrogen layer (Ib with H) or type IIb supernova (IIb), 
and type Ib supernova without hydrogen (Ib without H).  
Here we assumed $M_\mathrm{He} < 0.5~\mathrm{M_\odot}$ for SNe Ic
progenitors.  Note that the each boundary would shift to a higher value of
$M_\mathrm{ZAMS}$ for close binary systems with Case A mass transfer.
}\label{fig:mimf} 
\end{figure}

\begin{figure}
\epsscale{1.}
\plotone{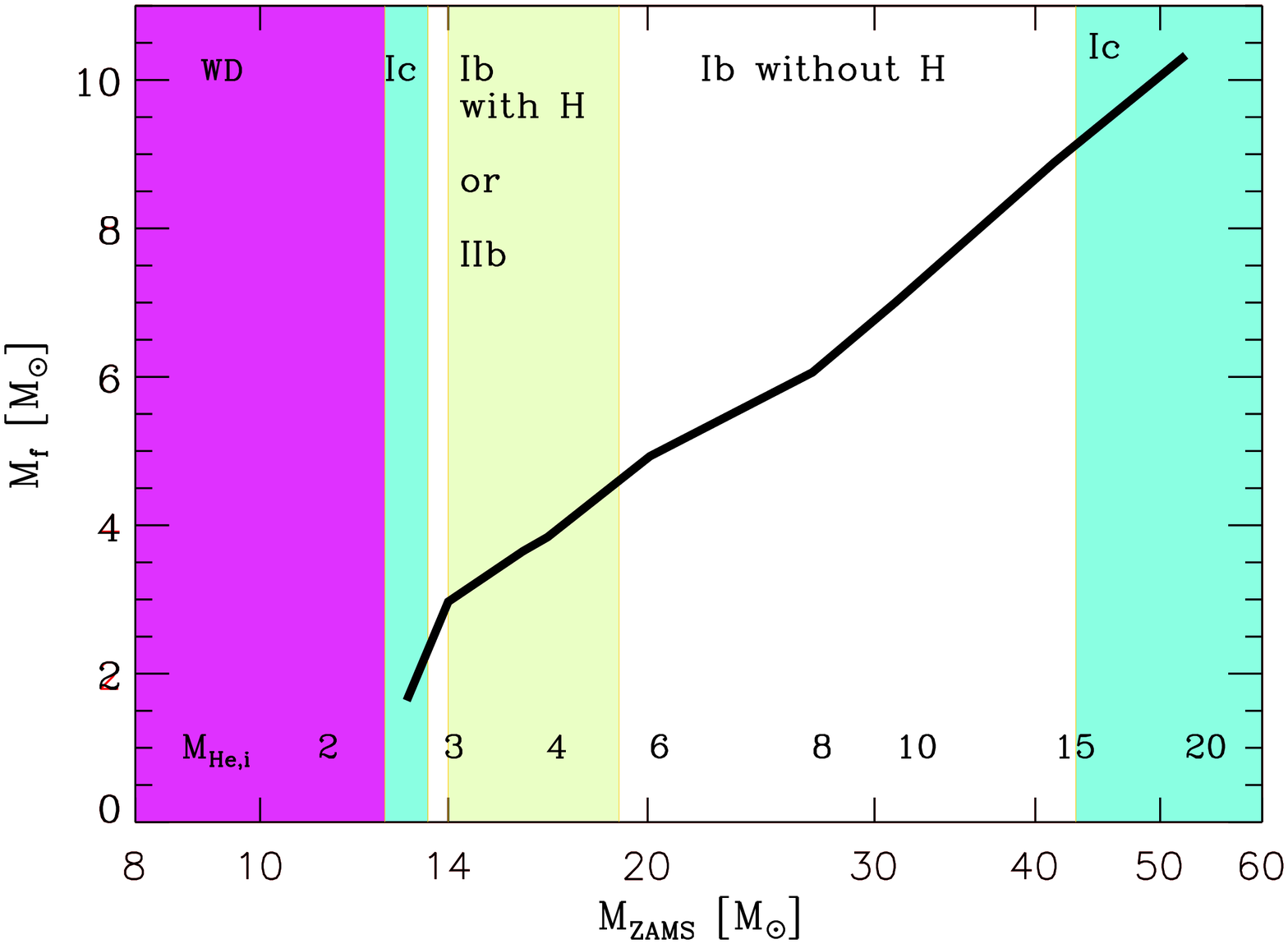}
\caption{
Same as in Fig.~\ref{fig:mimf}, but with $f_\mathrm{WR}=10$.}
\label{fig:mimf2} 
\end{figure}

Although the weak signature or no evidence of helium in SNe Ic spectra may
indicate the deficiency of helium in their progenitors, it is not well known how
much helium can be hidden in the supernova spectra.  It may also depend on the
degree of mixing of nickel into helium rich layers \citep{WE95}.  If we assume
0.5~\Msun{}, for instance, as the maximum amount of helium allowed for hiding
helium lines in SN spectra, our models indicate that most SNe Ic progenitors at
solar metallicity should belong to two distinct classes in terms of both ZAMS and
final masses, as summarized in Fig.~\ref{fig:mimf} for the binary systems that
undergo Case B mass transfer\footnote{For the systems with Case A mass
transfer, the each boundary should move to the right in the figure, but the
parameter space explored with our model grid is not large enough to determine
it quantitatively}. The same conclusion was also drawn by WL99 and
\citet{Pols02}. But Pols \& Dewi considered different types of binary systems
(see below), and the finding of "two mass classes" for the final masses  was
not obvious in WL99 due to the very high WR rate adopted in their study,
although it was clearly seen for the ZAMS masses.

If we assume that primary stars of $12.5 \la M_\mathrm{ZAMS}/\mathrm{M_\odot}
\la 13.5$ produce low-mass-class SNe Ic via Case BB mass transfer, about 62\%
of the SNe Ic from Case B systems should belong to the high mass class and the
rest ($\sim$38\%) to the low mass class, at $Z \approx \mathrm{Z_\odot}$ (see
Fig.~\ref{fig:mimf}).  It is important to note that the two classes are
produced by different mechanisms.  The high mass class of SNe Ic progenitors
(i.e., $M_\mathrm{ZAMS} \ga 33$~\Msun{} in Fig~\ref{fig:mimf}) is a consequence
of WR winds mass loss, while the low mass class results from Case ABB/BB mass
transfer  as discussed in Sect.~\ref{sect:masses}.  The ZAMS mass
range for the low mass class may not be much affected by metallicity, while it
should be widened with increasing metallicity for the high mass class.  This
leads to the conclusion that the low and high mass classes would dominate at
low and high metallicities respectively. It should also be noted that the final
mass range for the low mass class may not change much for different
metallicities, while it may decrease with increasing metallicity for the high
mass class, due to the increasing WR winds mass loss rates, as implied by the
result of WL99. 

In the massive close binary systems considered in this paper, the primary star
masses become much smaller than those of the secondary stars when Case ABB or
Case BB mass transfer begins. If the companion star mass were lower than the
helium star in a close binary system, the mass transfer rate should become
higher than in the systems of the present study.  For example, if a helium star
is located in a very short period binary system ($P \la 1~\mathrm{day}$) with a
less massive companion (e.g. a neutron star),  mass transfer from the helium
star may occur rapidly enough to make a helium-deficient carbon star, even for
$M_\mathrm{He,i} \approx$ 6.0~\Msun{} as shown by \citet{Pols02},
\citet{Dewi02} and \citet{Ivanova03}.  The final masses of such SN Ic
progenitors may range from 1.5~\Msun{} to 3.0~\Msun{}.  This scenario was also
suggested by \citet{Nomoto94} to explain the fast light curve of Type Ic SN
1994I.  The ZAMS mass of such a SN Ic should be in the range of 12 --
20~\Msun{} \citep{Pols02}. However, such close helium star plus neutron star systems
are supposed to rarely form, and might not contribute much to the population
of SNe Ic, compared to the systems considered in this study. 

Our results should have several observational consequences.  As mentioned
above,  the population of SNe Ic should be dominated by the high mass class for
$Z \ga \mathrm{Z_\odot}$. They would have higher ZAMS and final masses than
those of typical SNe Ib progenitors \citep[cf.][]{Kelly08, Anderson08}. Given
that the parameter space for the high mass class SNe Ic may become larger with
a higher WR mass loss rate, the number ratio of SNe Ic to SNe Ib, and that of
high mass class SNe Ic to low mass class SNe Ic should increase with increasing
metallicity \citep[cf.][]{Prieto08, Anderson09, Boissier09}. The existence of
the two mass classes of SNe Ic progenitors may be related to some aspects of
the observational diversity of SNe Ic.  For example, SNe Ic of the low mass
class is likely to be characterized by rather fast declining light curves and
low luminosities \citep[cf.][]{Iwamoto94, Richardson06, Young09},  implying
that the observed population of SNe Ic is likely to be baised to
high-mass-class SNe Ic. 

We should also note the huge difference of the binding energy between the two
classes.  The binding energy of the envelope above 1.4~\Msun{} in the SN Ic
progenitor star model of Seq.~3 ($M_\mathrm{f} = 1.64$~\Msun{}) is only about
$10^49$~erg, while it should be one or two orders of magnitude higher for a SN
Ic progenitor with $M_\mathrm{f} \ga 5.5$~\Msun{} (see Table~\ref{tab2}).
As a consequence, the energetics of SNe Ic might be systematically different for the
two different classes.

On the other hand, the assumption of $M_\mathrm{He} < 0.5$~\Msun{} for SN Ic
progenitors leads to a ratio of type Ic to Ib supernova rate (Ic/Ib ratio) of
about 0.4 from binary systems at solar metallicity 
\footnote{ At SMC
metallicity, the SN Ibc progenitor models of initial masses of 16 -- 40~\Msun{}
have $3.9\la M_\mathrm{f} \la 12$~\Msun{} with $M_\mathrm{He} \ga 1.2$~\Msun{}
(see Table~\ref{tab1}).  This implies that the low-mass class SNe Ic would
predominantly occur in binary systems at this metallicity.  If we assume stars
with $12.5 \la M_\mathrm{init} \la 13.5$ would produce low-mass class SNe Ic as
in the case of solar metallicity,  the SN Ic/SN Ib ratio would be about 0.1 at SMC
metallicity.
The exploration of the exact mass range for
the low-mass class SNe Ic is a time-consuming task, and we plan to investigate
this in near future. 
}.  
This appears in contradiction with recent observations that
indicate rather a high Ic/Ib ratio of about 2.0~\citep[e.g.][]{Smartt09}.  This
discrepancy would become even larger with $f_\mathrm{WR} =10$, as implied by
Fig.~\ref{fig:mimf2}. This raises a question on the nature of SN Ic progenitors,   
and it should be kept in mind that we still do not
fully understand what distinguishes SN Ic progenitors from those of SN Ib.  

A recent work by \citet{Dessart10} indicates that the mass fraction of helium
in the outermost layers ($Y_\mathrm{s}$), rather than the total mass of helium,
may be more relevant for the presence of helium lines in supernova spectra.
Specifically, it is shown that if helium is well mixed with CO material such
that $Y_\mathrm{s}$ becomes less than about 0.5, helium lines are not seen in
early time spectra, despite rather a large total amount of helium
($M_\mathrm{He} \simeq 1.0$~\Msun), if non-thermal excitation is absent.  
In our progenitor models, such a small
$Y_\mathrm{s}$ is realized only for $M_\mathrm{f} \ga$ 5.5~\Msun{} at solar
metallicity (with $f_\mathrm{WR} = 5$).  This is not different from the
above-discussed mass limit for having $M_\mathrm{He} \la 0.5$~\Msun, implying
that the initial mass range for SN Ic progenitors would not change much even if
we adopted $Y_\mathrm{s}$ as a criterion, at least for the high mass class. On
the other hand, we have $Y_\mathrm{s} = 0.98$ in the primary star of Seq.~3 at
carbon exhaustion while the total mass of helium is less than 0.2~\Msun. This
imples that the initial mass range for the low mass class SN Ic progenitors
might be affected if the condition of $Y_\mathrm{s} < 0.5$ for SN Ic
progenitors were applied.  But \citet{Dessart10} did not yet calculate such a
low mass SN progenitor model ($M_\mathrm{f} \la 2.0$~\Msun), and their analyses
were limited to early times of supernovae.  
It remains to be an important subject of
future work to systematically investigate which types of supernova progenitors
would lead to the presence or absence of helium lines
in the supernova spectra at different epochs, including the effect of 
non-thermal excitation. 
Therefore, the above discussion on Type Ic progenitors based on 
the total amount of helium should only be considered indicative at this stage.

\subsection{Presence of hydrogen}\label{sect:dischydrogen}

It is interesting that, at $Z \approx \mathrm{Z_\odot}$, the presence of a thin
hydrogen layer is only expected for a limited range of the initial/final mass
of SN Ib progenitors, as shown in Figs~\ref{fig:dmh}, and~\ref{fig:mimf}. 
The detection of hydrogen absorption lines at high velocity
has been indeed reported in many SNe Ib \citep[e.g.,][]{Deng00, Branch02,
Elmhamdi06},  in favor of our model prediction for the presence of a thin
hydrogen layer in SNe Ib progenitors.  This might provide a strong constraint
for the progenitor masses of observed SNe Ib, in principle.  

Note also that explosions of such helium stars with thin hydyrogen layers could
be recognized as SN IIb rather than Ib, if hydrogen lines were detected short
after supernova explosion, e.g., as in the case of SN 2008ax~\citep{Chornock10}
and as recently discussed by \citet{Baron10}   and \citet{Dessart10}.
The radii of these progenitor models range from $\sim 10^{11}~\mathrm{cm}$ to
$\sim 10^{12}~\mathrm{cm}$. They may corredpond to the ''compact'' category of
SN IIb progenitors, which is discussed in \citet{Chevalier10}. The relatively
low ejecta masses of such SNe IIb are consitent with our model predictions.

On the other hand, Case~C mass transfer can also leave helium cores covered  with
small amounts of hydrogen envelope. As the life time of such stars made via Case
C mass transfer should be rather short, they can retain much more
hydrogen ($M_\mathrm{H} > 0.1$~\Msun), than what is predicted from our binary
models with Case AB/B mass transfer. Such a star may eventually explode as a SN
IIb like SN 1993J \citep[e.g.][]{Podsiadlowski93, Maund04}, with a much
extended envelope ($\sim 10^{13} - 10^{14}~\mathrm{cm}$). 

Therefore, the two categories of SN IIb progenitors according to their sizes,
which has been recently suggested by \citet{Chevalier10}, may be understood
within the framework of binary evolution; SNe IIb of the compact type  may be
produced via Case AB/B mass transfer (especially at $Z \la \mathrm{Z_\odot}$),
and SNe IIb of the extended type via Case C mass transfer.  

\acknowledgments
This work is supported
by the DOE SciDAC Program (DOE DE-FC02-06ER41438), the NSF grant (NSF-ARRA
AST-0909129), and the NASA Theory Program (NNX09AK36G). 
We are grateful to Luc Dessart for useful discussions.

\clearpage

\begin{center}

\begin{deluxetable}{cccccccccccccc}
\tabletypesize{\scriptsize}
\tablewidth{0pt}
\tablecaption{Properties of the computed sequences\tablenotemark{~} \label{tab1}}
\tablehead{
\colhead{No.} & \colhead{$Z$} & \colhead{$M_\mathrm{1,i}$} & \colhead{$M_\mathrm{2,i}$} & \colhead{$P_\mathrm{i}$} & \colhead{$f_\mathrm{WR}$} &  %
\colhead{Case} & \colhead{$P_\mathrm{f}$} &  \colhead{$M_\mathrm{1,f}$} & \colhead{$M_\mathrm{CO,f}$} & \colhead{$M_\mathrm{He}$} & \colhead{$M_\mathrm{H}$}  %
&  \colhead{$<j_\mathrm{1.4}>$} & \colhead{Fate}
}
\startdata
1 & 0.02 & 12 & 8 & 3.0 & 5 &  B+BB & 57.9 &  1.40\tablenotemark{e}& 1.21\tablenotemark{e} & 0.17\tablenotemark{e}  & 0.0  &  0.33  & ONeMg WD  \\
2 & 0.02 & 12 & 11 & 4.0 & 5 & B+BB &  104.4 & 1.48 \tablenotemark{e}& 1.24\tablenotemark{e} & 0.20\tablenotemark{e} & 0.0 &  0.35  & ONeMg WD \\
3 & 0.02 & 13 & 11 & 5.0 & 5 & B+BB & 123.3 & 1.64\tablenotemark{e} & 1.43\tablenotemark{e} & 0.18\tablenotemark{e} & 0.0 &  0.22 & SN Ic\\
4 & 0.02 & 14 & 12 & 3.0 & 5 & A+AB+BB & 118.5 & 1.33\tablenotemark{e} & 1.09\tablenotemark{e} & 0.22\tablenotemark{e} & 0.0  &  $-$ & ONeMg WD \\ 
5 & 0.02 & 14 & 12 & 5.0 & 5 & B+BB &  30.7 & 2.97 & 1.66 & 1.24 & 1.9(-4) &  0.25 & SNIb \\ 
6 & 0.02 & 16 & 14 & 2.0 & 5 &  A:Contact &  &      &   &  &  & &    \\
7 & 0.02 & 16 & 14 & 3.0 & 5 &  A+AB+ABB & 101.8  & 1.54\tablenotemark{e} & 1.33\tablenotemark{e} & 0.17\tablenotemark{e}  & 0.0&  0.39  & ONeMg WD \\
8 & 0.02 & 16 & 14 & 4.0 & 5 & B+BB & 26.2 & 3.66 & 2.05 & 1.47 & 4.5(-3) &  0.24 & SNIb \\ 
9 & 0.02 & 16 & 14 & 5.0 & 5 &  B+BB & 33.7 & 3.65 &  2.04  & 1.47  & $5.0(-3)$ &  0.25 &  SNIb  \\
10 & 0.02 & 18 & 12 & 3.0 & 5 &  A+AB+ABB & 27.9 & 2.66 &  1.58  & 1.01  & $0.00$ & 0.26 & SNIb  \\
11 & 0.02 & 18 & 12 & 3.0 & 10 &  A+AB+ABB & 27.3 & 2.74 &  1.59  & 1.08  & $0.00$ &  0.26 & SNIb  \\
12 & 0.02 & 18 & 12 & 5.0 & 10 &  B:Contact &  &  &    &   &  &    &  \\
13 & 0.02 & 18 & 17 & 3.0 & 10 &  A+AB+ABB  & 36.2 & 3.03  &  1.68   & 1.27 & $7.9(-4)$ &  0.25 & SN Ib   \\
14 & 0.02 & 18 & 17 & 4.0 & 10 &  A+AB & 29.7 & 3.79  &  2.14   & 1.49 & $1.5(-3)$ &  0.25 & SN Ib   \\
15\tablenotemark{a}  & 0.02 & 18 & 17 & 4.0 & 10 &  A+AB & 24.0 & 3.97  &  2.27   & 1.53 & $1.0(-3)$ &  0.26 & SNIb   \\
16\tablenotemark{b}  & 0.02 & 18 & 17 & 4.0 & 10 &  A+AB & 50.0 & 3.80  &  2.16   & 1.50 & $2.2(-3)$ &  0.06 & SNIb   \\
17\tablenotemark{c}  & 0.02 & 18 & 17 & 4.0 & 10 &  A+AB & 25.2 & 3.84  &  2.18   & 1.50 & $1.4(-3)$  & 0.26 & SNIb   \\
18\tablenotemark{d}  & 0.02 & 18 & 17 & 4.0 & 10 &  A+AB & 30.6 & 3.73  &  2.14   & 1.43 & $ 0.00 $ &  3.57 & SNIb  \\
19  & 0.02 & 18 & 17 & 5.0 &  3 &  B    & 33.1 & 3.73  &  2.33   & 1.23 & $0.00$ & 0.25 & SNIb   \\
20  & 0.02 & 18 & 17 & 5.0 &  5 &  B   & 32.4 & 4.04  & 2.45  & 1.4 & 0.00 &  0.33 & SNIb \\ 
21  & 0.02 & 18 & 17 & 5.0 & 10 &  B+BB & 31.5 & 4.41  &  2.51   & 1.68 & $9.9(-3)$ & 0.26 & SNIb   \\
22  & 0.02 & 18 & 17 & 6.0 & 10 &  B    & 39.3 & 4.39  &  2.56   & 1.62 & $4.0(-3)$ & 0.26 & SNIb   \\
23  & 0.02 & 25 & 19 & 6.0 & 10 & B:Contact  &  &   &     &  &  & &      \\
24  & 0.02 & 25 & 24 & 2.0 & 10 & A:Contact  &  &   &     &  &  & &     \\
25  & 0.02 & 25 & 24 & 3.0 & 3 & A+AB  & 22.7  & 3.70  & 2.46  &0.98  & 0.0  & 0.24 & SNIb  \\
26  & 0.02 & 25 & 24 & 3.0 & 5 & A+AB  & 22.4 & 4.33  & 2.80 & 1.30  & 0.0  &  0.25 & SNIb  \\
27  & 0.02 & 25 & 24 & 3.0 & 10 & A+AB & 21.3  & 5.07  & 3.17  &1.67  & 0.0  & 0.25 & SNIb   \\
28\tablenotemark{c}  & 0.02 & 25 & 24 & 3.0 & 10 & A+AB & 18.9 & 5.08  & 3.19  &1.66  & 0.0  & 0.32 & SNIb   \\
29  & 0.02 & 25 & 24 & 4.0 & 5 & A + AB & 21.5 & 4.45 & 2.91 & 1.22 & 0.0 &  0.26 & SNIb \\
30  & 0.02 & 25 & 24 & 6.0 & 10 & B & 27.4  & 6.49 & 4.45  & 1.63 & 0.0 &  0.39 & SNIb   \\
31  & 0.02 & 60 & 40 & 7.0 & 3 & A & 16.8   & 4.95  & 3.70 & 0.25  & 0  &  0.24 &  SNIc  \\
\hline
32  & 0.004 & 16 & 12 & 3.0 & 5 & B+BB & 64.75 & 3.91  & 2.22 & 1.54  & $1.6(-2)$  &   0.24 &  SNIb  \\
33  & 0.004 & 16 & 14 & 3.0 & 5 & B+BB & 19.6  & 3.90  & 2.21 & 1.53  & $1.6(-2)$  &   0.24 &  SNIb  \\
34  & 0.004 & 16 & 14 & 5.0 & 5 & B+BB & 21.8  & 3.84  & 2.19 & 1.51  & $1.2(-2)$  &  0.24 & SNIb \\
35  & 0.004 & 18 & 12 & 5.0 & 5 & B+BB & 14.4 & 4.64 & 2.76  & 1.68  & $1.7(-2)$ &   0.31 & SNIb  \\
36  & 0.004 & 18 & 12 & 8.0 & 5 & B+BB & 24.33 & 4.56 & 2.68 & 1.67  & $1.5(-2)$ &  0.33 & SNIb \\
37  & 0.004 & 18 & 17 & 3.0 & 5 & A+AB & 18.4   & 4.42  & 2.55 & 1.67  & $2.6(-2)$  & 0.26 &  SNIb  \\
38  & 0.004 & 18 & 17 & 3.0 & 10 & A+AB &14.4 & 4.58  & 2.61 & 1.77  & $2.7(-2)$  & 0.26 &  SNIb  \\
39  & 0.004 & 18 & 17 & 6.0 & 5 & B+BB & 27.0 & 4.57 & 2.71  &  1.65 & $1.6(-2)$ &  0.27 & SNIb   \\
40  & 0.004 & 25 & 12 & 3.0 & 5 & A: Contact &  &  &    &   & & &     \\
41  & 0.004 & 25 & 12 & 6.0 & 5 & B: Contact  &  &  &    &   & & &      \\
42  & 0.004 & 25 & 19 & 3.0 & 5 & A+AB & 10.2 & 7.09 & 4.87   & 2.03  &$6.5(-3)$ & 0.32& SNIb \\
43  & 0.004 & 25 & 24 & 3.0 & 5 & A+AB & 13.0 & 7.31 & 5.05  & 2.07  &$8.7(-3)$ & 0.28 & SNIb   \\
44  & 0.004 & 25 & 24 & 6.0 & 5 & B:Contact &  &  &    &  & &  \\
45  & 0.004 & 40 & 30 & 4.0 & 5 & A+AB & 9.31  & 12.0  & 9.42 & 1.24  & $0.0$  &  0.56 &  BH  
\enddata
\tablenotetext{~}{Each column has the following meaning. Z: the adopted metallicity, 
$M_\mathrm{1,i}$: the initial mass of the primary component in units of \Msun,
$M_\mathrm{2,i}$: the initial mass of the secondary component in units of \Msun, 
$P_\mathrm{i}$: the initial orbital period in units of day,  
$f_\mathrm{WR}$: the correction factor for the WR winds mass rate given in Eq.~(\ref{eq1}), 
Case: the mass transfer case, 
$P_\mathrm{f}$: the orbital period in units of day, at the end of calculation (mostly at neon burning in the primary star), 
$M_\mathrm{1,f}$: the mass of the primary star at the end of calculation in units of \Msun, 
$M_\mathrm{CO, f}$: the CO core mass of the primary star at the end of calculation in  units of \Msun, 
$M_\mathrm{He}$: the helium mass in the envelope of the primary star at the end of calculation in units of \Msun, 
$M_\mathrm{H}$: the hydrogen mass in the envelope of the primary star at the end of calculation in units of \Msun, 
$<j_\mathrm{1.4}>$: the specific angular momentum in the innermost 1.4~$\mathrm{M_\odot}$ of the primary star at the end of calculation 
in  units of $10^{15} \mathrm{cm s^{-1}}$, 
Fate: the expected final fate of the primary star.}
\tablenotetext{a}{$f_\mathrm{sync}=10^5$}
\tablenotetext{b}{$f_\mathrm{sync}=0.01$}
\tablenotetext{c}{$\tau_\mathrm{sync}$ according to Zahn}
\tablenotetext{d}{Non-magnetic model}
\tablenotetext{e}{The values are lower limits for $M_\mathrm{1,f}$ and $M_\mathrm{CO,f}$, and upper limits for $M_\mathrm{He}$
since the calculation was stop long before the neon burning phase while the Case BB/ABB mass transfer phase was not finished.}
\end{deluxetable}

\end{center}

\clearpage

\begin{center}
\begin{deluxetable}{cccccccc}
\tablecaption{Properties of the computed single helium stars at $Z=0.02$\tablenotemark{~} \label{tab2}}
\tablewidth{0pt}
\tablehead{
 \colhead{$f_\mathrm{WR}$}  & \colhead{$M_\mathrm{He, i}$}  & \colhead{$M_\mathrm{f}$}  &{$M_\mathrm{CO,f}$} & \colhead{$ M_\mathrm{He}$} & \colhead{$R$} & \colhead{$R_\mathrm{95}$}& \colhead{$E_b$}} 
\startdata
5 & 2.8   & 2.73 &  1.51 & 1.16 & 163.54 & 5.08 & 0.01 \\ 
5 &  3.0  & 2.91 &  1.60 & 1.22 & 33.40 &  1.79 & 0.03 \\
5 &  4.0  & 3.72 &  2.15 & 1.43 & 8.01 & 0.73  & 0.18  \\ 
5 &   6.0  & 4.23 &  2.77 & 1.19 & 4.19 & 0.38 & 0.25  \\
5 &   8.0  & 4.95 &  3.40 & 0.85 & 3.07 & 0.30 & 0.29 \\ 
5 &  10.0  & 5.49 &  3.93 & 0.54 & 1.59 & 0.22 & 0.47 \\
5 &  15.0  & 6.44 &  4.85 & 0.31 & 0.59 & 0.18 & 0.55 \\
5 &  20.0  & 7.09 &  5.44 & 0.22 & 0.48 & 0.16 & 0.66 \\
\hline
10 &  3.0  &  2.95 & 1.60 & 1.27 & 34.96 &  1.84 & 0.03 \\
10 &  4.0  &  3.84 & 2.20 & 1.49 & 6.94  &  0.70 & 0.15 \\
10 &  6.0  &  4.93 & 3.14 & 1.55 & 4.29 & 0.51  & 0.19  \\
10 &  8.0  &  6.06 & 4.13 & 1.41 & 2.55  & 0.35  & 0.33  \\ 
10 & 10.0  &  7.01 & 4.96 & 1.12 & 1.94 & 0.25 & 0.55   \\
10 & 15.0  & 8.89 & 6.80 &  0.47 & 0.54 & 0.20 & 0.82 \\
10 & 20.0  & 10.33 & 8.21 & 0.31 & 0.40 & 0.16  & 1.17   
\enddata
\tablenotetext{~}{Each column has the following meaning. 
$f_\mathrm{WR}$: the correction factor for the WR winds mass rate given in Eq.~(\ref{eq1}), 
$M_\mathrm{He,i}$: the initial mass of the helium star in units of \Msun, 
$M_\mathrm{f}$: the final mass at the end of calculation (i.e., at neon burning) in units of \Msun, 
$M_\mathrm{CO,f}$: the CO core mass at the end of calculation in units of \Msun,
$M_\mathrm{He}$: the amount of Helium in the envelope at the end of calculation in units of \Msun,
$R$: the radius at the end of calculation in units of $\mathrm{R_\odot}$,
$R_\mathrm{95}$: the radius that encompasses 95 \% of the total mass at the end of calculation in units of $\mathrm{R_\odot}$, 
$E_\mathrm{b}$: the binding energy of the envelope above $1.5~\mathrm{M_\odot}$ at the end of calculation in units of $10^{51} \mathrm{erg}$}
\end{deluxetable}
\end{center}

\clearpage

\end{document}